\documentclass[sigconf,nonacm]{acmart}
\AtBeginDocument{%
  }

\usepackage[camera]{dtrt-acm}
\usepackage{tikz}
\usepackage{amsmath}

\usepackage{amsmath}
\usepackage{amsthm}
\usepackage{booktabs}
\usepackage{enumitem}
\usepackage{mathpartir}
\usepackage{multirow}
\usepackage{pifont}
\usepackage{stmaryrd}
\usepackage{substr}
\usepackage{xspace}
\usepackage{xcolor}

\makeatletter
\@namedef{T1/zi4/m/it}{<->ssub*zi4/m/n}
\@namedef{T1/zi4/b/it}{<->ssub*zi4/b/n}
\@namedef{T1/zi4/bx/it}{<->ssub*zi4/b/n}
\@namedef{T1/zi4/m/sl}{<->ssub*zi4/m/n}
\@namedef{T1/zi4/b/sl}{<->ssub*zi4/b/n}
\@namedef{T1/zi4/bx/sl}{<->ssub*zi4/b/n}
\@namedef{TS1/zi4/m/it}{<->ssub*zi4/m/n}
\@namedef{TS1/zi4/b/it}{<->ssub*zi4/b/n}
\@namedef{TS1/zi4/bx/it}{<->ssub*zi4/b/n}
\@namedef{TS1/zi4/m/sl}{<->ssub*zi4/m/n}
\@namedef{TS1/zi4/b/sl}{<->ssub*zi4/b/n}
\@namedef{TS1/zi4/bx/sl}{<->ssub*zi4/b/n}
\makeatother

\newcommand{\revision}[1]{#1}

\newcommand{\name}{\textsc{skanf}\xspace}
\newcommand{\switchtable}{branch table\xspace}
\newcommand{\numberofcontracts}{6,554\xspace}
\newcommand{\identifiedbugs}{1,046\xspace}
\newcommand{\confirmedbugs}{394\xspace}
\newcommand{\totalloss}{\$10.6M\xspace}
\newcommand{\contractcall}{external call}

\title{Insecurity Through Obscurity:\\Veiled Vulnerabilities in Closed-Source Contracts}
\renewcommand{\shorttitle}{Insecurity Through Obscurity: Veiled Vulnerabilities in Closed-Source Contracts}
\usepackage{xcolor}
\usepackage{listings}
\usepackage[most]{tcolorbox}

\definecolor{keywordcolor}{RGB}{197, 134, 192}  %
\definecolor{functioncolor}{RGB}{130, 170, 255} %
\definecolor{numbercolor}{RGB}{237, 151, 101}   %
\definecolor{tealgreen}{RGB}{32, 178, 170}    %

\lstdefinelanguage{YulAssembly}{
    keywords={jumpi, jump, eq, origin, revert, stop, calldataload, shr, call, iszero, mstore, mload, sub, add},
    keywordstyle=\color{blue}\bfseries,
    morekeywords={let, if, jumpdest, else},
    morecomment=[l]{//},   
    morestring=[b]",
    moredelim=[s][\color{tealgreen}]{0x}{\ },
}

\lstdefinestyle{YulStyle}{
    language=YulAssembly,
    basicstyle=\small\ttfamily,
    commentstyle=\itshape\color{gray},
    stringstyle=\color{stringcolor},
    columns=fullflexible,
    frame=none,
    rulecolor=\color{gray},
    breaklines=true,
    numbers=left,
}

\definecolor{lightgray}{rgb}{0.95,0.95,0.95}  

\lstdefinelanguage{EVMAssembly}{
    keywords={DUP1, PUSH2, EQ, JUMPI, JUMPDEST},
    keywordstyle=\color{blue}\bfseries,
    morecomment=[l]{//},   
    morestring=[b]",
    moredelim=[s][\color{tealgreen}]{0x}{\ },
}

\lstdefinestyle{EVMStyle}{
    language=EVMAssembly,
    basicstyle=\small\ttfamily,  
    commentstyle=\itshape\color{gray},  
    stringstyle=\color{gray},    
    columns=fullflexible,
    frame=single, 
    framerule=0pt,
    rulecolor=\color{gray},
    breaklines=true,
    numbers=none,
    numberstyle=\color{gray},
    backgroundcolor=\color{lightgray},
}

\lstdefinelanguage{Solidity}{
	keywords=[1]{anonymous, assembly, assert, balance, break, call, callcode, case, catch, class, constant, continue, constructor, contract, debugger, default, delegatecall, delete, do, else, emit, event, experimental, export, external, false, finally, for, function, gas, if, else, implements, import, in, indexed, instanceof, interface, internal, is, length, library, log0, log1, log2, log3, log4, memory, modifier, new, payable, pragma, private, protected, public, pure, push, require, return, returns, revert, selfdestruct, send, solidity, storage, struct, suicide, super, switch, then, this, throw, transfer, true, try, typeof, using, value, view, while, with, addmod, ecrecover, keccak256, mulmod, ripemd160, sha256, sha3, fallback}, %
	keywordstyle=[1]\color{blue}\bfseries,
	keywords=[2]{address, bool, byte, bytes, bytes1, bytes2, bytes3, bytes4, bytes5, bytes6, bytes7, bytes8, bytes9, bytes10, bytes11, bytes12, bytes13, bytes14, bytes15, bytes16, bytes17, bytes18, bytes19, bytes20, bytes21, bytes22, bytes23, bytes24, bytes25, bytes26, bytes27, bytes28, bytes29, bytes30, bytes31, bytes32, enum, int, int8, int16, int24, int32, int40, int48, int56, int64, int72, int80, int88, int96, int104, int112, int120, int128, int136, int144, int152, int160, int168, int176, int184, int192, int200, int208, int216, int224, int232, int240, int248, int256, mapping, string, uint, uint8, uint16, uint24, uint32, uint40, uint48, uint56, uint64, uint72, uint80, uint88, uint96, uint104, uint112, uint120, uint128, uint136, uint144, uint152, uint160, uint168, uint176, uint184, uint192, uint200, uint208, uint216, uint224, uint232, uint240, uint248, uint256, var, void, ether, finney, szabo, wei, days, hours, minutes, seconds, weeks, years},	%
	keywordstyle=[2]\color{teal}\bfseries,
	keywords=[3]{block, blockhash, coinbase, difficulty, gaslimit, number, timestamp, msg, data, gas, sender, sig, value, now, tx, gasprice, origin},	%
	keywordstyle=[3]\color{violet}\bfseries,
	identifierstyle=\color{black},
	sensitive=true,
	comment=[l]{//},
	morecomment=[s]{/*}{*/},
	commentstyle=\color{gray}\ttfamily,
	stringstyle=\color{red}\ttfamily,
	morestring=[b]',
	morestring=[b]"
}

\lstdefinestyle{SolidityStyle}{
    language=Solidity,
    basicstyle=\small\ttfamily,
    commentstyle=\itshape\color{gray},
    stringstyle=\color{stringcolor},
    columns=fullflexible,
    frame=none,
    rulecolor=\color{gray},
    breaklines=true,
    numbers=left,
}

\lstdefinelanguage{json}{basicstyle=\small\ttfamily,  
    commentstyle=\itshape\color{gray},  
    stringstyle=\color{gray},    
    columns=fullflexible,
    morecomment=[l]{//},  
    frame=single, 
    framerule=0pt,
    rulecolor=\color{gray},
    breaklines=true,
    numberstyle=\color{gray},
    numbers=left,
stringstyle=\color{brown}\ttfamily,
    morestring=[b]',
    morestring=[b]",
    literate=
  *{0}{{{\color{teal}0}}}{1}
   {1}{{{\color{teal}1}}}{1}
   {2}{{{\color{teal}2}}}{1}
   {3}{{{\color{teal}3}}}{1}
   {4}{{{\color{teal}4}}}{1}
   {5}{{{\color{teal}5}}}{1}
   {6}{{{\color{teal}6}}}{1}
   {7}{{{\color{teal}7}}}{1}
   {8}{{{\color{teal}8}}}{1}
   {9}{{{\color{teal}9}}}{1}
   {"caller":}{{{\color{blue}"caller"}}:}1
{"origin":}{{{\color{blue}"origin"}}:}1
{"blockNumber":}{{{\color{blue}"blockNumber"}}:}1
{"callPC":}{{{\color{blue}"callPC"}}:}1
{"calldata":}{{{\color{blue}"calldata"}}:}1
{"targetAddress":}{{{\color{blue}"targetAddress"}}:}1
{"functionSelector":}{{{\color{blue}"functionSelector"}}:}1
{"destination":}{{{\color{blue}"destination"}}:}1
{"amount":}{{{\color{blue}"amount"}}:}1
}

\begin{document}
\author[Sen Yang]{Sen Yang}
\orcid{0000-0002-8866-2097}
\affiliation{%
  \institution{Yale University, IC3}
  \city{New Haven}
  \state{Connecticut}
  \country{United States}
}
\email{sen.yang@yale.edu}

\author[Kaihua Qin]{Kaihua Qin}
\orcid{0000-0003-2190-3623}
\authornote{Part of this work was conducted while the author was affiliated with Yale University.}
\affiliation{
    \institution{University of Warwick}
    \city{Coventry}
    \state{West Midlands}
    \country{United Kingdom}
}
\email{kaihua@qin.ac}

\author[Aviv Yaish]{Aviv Yaish}
\orcid{0000-0002-7971-2494}
\affiliation{%
  \institution{Yale University, IC3}
  \city{New Haven}
  \state{Connecticut}
  \country{United States}
}
\email{a@yai.sh}

\author[Fan Zhang]{Fan Zhang}
\orcid{0000-0002-8525-4514}
\affiliation{%
  \institution{Yale University, IC3}
  \city{New Haven}
  \state{Connecticut}
  \country{United States}
}
\email{f.zhang@yale.edu}

\begin{abstract}

Most blockchains cannot hide the binary code of programs (i.e., smart contracts) running on them.
To conceal proprietary business logic and to potentially deter attacks,
many smart contracts are closed-source and in many cases exhibit code obfuscation, either intentionally introduced to hide internal logic or unintentionally produced by optimizations.
However, we demonstrate that such obfuscation can obscure critical vulnerabilities rather than enhance security, 
a phenomenon known as \textit{insecurity through obscurity}. 
To systematically analyze these risks on a large scale, we present \name, a novel EVM bytecode analysis tool tailored for closed-source and obfuscated contracts. \name combines control-flow deobfuscation with symbolic execution based on historical transactions to identify and exploit asset management vulnerabilities. 
Our evaluation on real-world Maximal Extractable Value (MEV) bots reveals that \name detects vulnerabilities in \identifiedbugs contracts and successfully generates exploits for \confirmedbugs of them, with potential losses of \$10.6M. Additionally, we uncover 104 real-world MEV bot attacks that collectively resulted in \$2.76M in losses. 

\end{abstract}



\maketitle

\section{Introduction}

On public blockchains such as Ethereum, smart contract bytecode---and source code, if developers choose to release it---are publicly accessible.
While transparency is a core feature, it allows a contract's logic to be examined, reverse-engineered, or copied.
As a result, many developers choose not to publish their smart contract source code.
Furthermore, {\em bytecode obfuscation} is commonly observed, either intentionally introduced by developers to protect their contracts from attacks~\cite{degatchi2023smart,degatchi2024mev,ma2025surviving}, or as a byproduct of compiler optimizations in languages such as Vyper~\cite{vyper} and Huff~\cite{huff}.

To provide context, our empirical study of Ethereum  (\autoref{sec:ethereum-smart-contracts}) shows that about 15\% of the most active contracts (top 50K, ranked by the number of transactions interacted with them) are closed-source. 
Among these closed-source contracts, obfuscated contracts collectively hold more than \$37M worth of crypto assets.

While obfuscation may have certain benefits due to ``security through obscurity'' (e.g., some practitioners deem obfuscation necessary for deterring frontrunning attacks~\cite{degatchi2024mev}), this paper is motivated by the observation that obfuscation can also introduce \emph{insecurity}.

To see the motivation, consider the recent ``Destroyer Inu'' incident, where 22 ETH (worth \$51,056 at the time) were stolen from a closed-source smart contract equipped with layers of obfuscation.
The attack exploits a \emph{well-known} vulnerable pattern (misuse of tx.origin~\cite{soliditybyexample2025origin} followed by improper asset management), but existing tools could not detect the full exploit because of control flow obfuscation. E.g., Mythril~\cite{mythril} recognizes the superficial misuse of tx.origin but misses the much more severe asset management vulnerability, only reporting a ``low'' risk.
In this case, obscurity arguably renders the contract \emph{less} secure by hindering analysis, making this a prime instance of insecurity through obscurity \cite{meunier2008cwe}\footnote{Compare this with modern cryptographic schemes following Kerckhoffs's principle that a system should not be compromised if adversaries uncover its method of operation \cite{petitcolas2019kerckhoffs}, or, as Shannon said: \emph{``the enemy knows the system''} \cite{shannon1949communication}.}.

\parhead{This work}
While bytecode-level analyzers exist, they remain limited in their ability to scale to large, closed-source contracts, especially when control-flow obfuscation is present, leaving gaps in our understanding of their security, particularly for asset-heavy targets attractive to attackers. In this paper, we aim to develop effective code analysis techniques to address this gap.
We focus on a broad class of security vulnerabilities that we refer to as {\em asset management vulnerabilities}. These vulnerabilities allow an adversary to steal assets, creating a strong incentive for exploitation if the vulnerabilities cannot be detected promptly. 

\parhead{Real-world testbed: MEV bots}
While our tool is general, we use MEV (Maximal Extractable Value) bots as the testbed for efficacy evaluation and as concrete examples to study the security of closed-source smart contracts in the wild.

MEV bots are highly optimized smart contracts deployed by MEV searchers to facilitate MEV extraction~\cite{daian2020flash}.
While some MEV bots are considered toxic, many play a beneficial role~\cite{ethereum2025mev,garimidi2025mev}. For example, arbitrage bots align DEX prices with the broader market, providing users with accurate prices.
We choose MEV bots as our testbed because they are commonly closed-source, and a subset of them are obfuscated (we identify \numberofcontracts closed-source MEV bots as of the time of writing; six of the top 10, ranked by the number of transaction bundles they sent~\cite{LibMEVLeaderboard2025}, exhibit some form of obfuscation).
Moreover, they handle large volumes of assets, making any asset management vulnerabilities particularly relevant and detrimental.

\smallskip\noindent 
We summarize our goals as three research questions (RQs):
\begin{itemize}
    \item \textbf{RQ1:} How can we effectively {\em deobfuscate} the control flow given smart contract bytecode? Vulnerability detection is only possible if this can be accomplished.
    \item \textbf{RQ2:} How can we scale vulnerability detection to complex closed-source contracts, where existing symbolic execution tools suffer from path explosion?
    Even after deobfuscation, detecting such vulnerabilities can still be challenging, given the complexity of MEV bots. 
    \item \textbf{RQ3:} How many MEV bot contracts have been exploited in practice, and how much did they lose? Answering this question helps in understanding the severity of asset management vulnerabilities and estimating how much loss could be reduced if \name were used by searchers.
\end{itemize}

\parhead{Challenges}
Numerous tools~\cite{dedaub,mythril,frank2020ethbmc,gritti2023confusum,grech2019gigahorse} are available to identify smart contract bugs. 
However, we highlight three limitations that render them inefficient for our tasks.

The first challenge is \textit{control-flow obfuscation}. Smart contracts may employ indirect jumps that depend on runtime values, obscuring the execution path. Existing tools typically misclassify such basic blocks as unreachable~\cite{grech2019gigahorse,grech2022elipmoc,lagouvardos2024incredible} or raise errors when encountering symbolic jump destinations~\cite{mythril,frank2020ethbmc,gritti2023confusum,ruaro2024not}.

Secondly, detecting asset management vulnerabilities requires a fine-grained analysis of a smart contract's logic, and simple pattern matching, as done by existing tools, does not suffice.
E.g., JACKAL~\cite{gritti2023confusum} flags \texttt{CALL} instructions with a fixed function selector as safe, overlooking vulnerabilities where function parameters can be manipulated.
Mythril~\cite{mythril} only flags vulnerable \texttt{CALL} instructions when calldata can be completely controlled by the adversary, missing attacks that only require partially adversarial calldata.

The third challenge is the \textit{complexity} of our real-world targets, which involve trading, flash loans, DEX swaps, and other interactions with DeFi protocols. This can result in path explosion and expensive constraint solving, making symbolic execution inefficient without scalable methods for input generation.

\parhead{Our Methods} {In this paper, we present the design and implementation of \name
\footnote{The name \name stands for \textbf{S}en, \textbf{K}aihua, \textbf{A}viv '\textbf{n} \textbf{F}an.}
, a tool for detecting asset management vulnerabilities in closed-source and obfuscated smart contracts.
\name addresses the above challenges with the following three key ideas.}

First, we de-obfuscate control flows by rebuilding jump tables. 
We observe that, unlike x86-64 or ARM64, where jumps can target any address, valid jump destinations in EVM (Ethereum Virtual Machine) must be marked with \texttt{JUMPDEST}~\cite{wood2014ethereum}. This enables us to reconstruct the control flow and instrument the bytecode with a \switchtable to improve the coverage of subsequent analyses.

Second, to detect asset management vulnerabilities, we combine symbolic execution and taint analysis to identify {\em vulnerable calls} that adversaries can trigger with malicious parameters, including those where only critical parameters are adversary-controlled.

Third, we use transaction-seeded symbolic execution, i.e., symbolic execution initialized with concrete inputs from historical transactions, to speed up symbolic execution and mitigate path explosion through concrete seed inputs~\cite{sen2007concolic}. While seeded symbolic execution itself is not new, our contribution lies in a smart-contract-specific realization that leverages public historical transactions as high-quality seeds. To obtain such seeds at a large scale, we observe that historical transactions, when available\done, can be leveraged to automatically extract input seeds to improve both efficiency and detection efficacy. 
In the absence of historical transactions, our symbolic execution still functions correctly, though without the efficiency gains from seed inputs.

\parhead{Responsible Disclosure}
As our exploits can readily compromise existing smart contracts, we responsibly disclosed them to the affected parties through a dedicated blockchain messaging service (Blockscan Chat by Etherscan~\cite{blockscan2025chat}), following the same disclosure practice adopted by prior research~\cite{qin2023blockchain}.
Moreover, we adhere to several additional responsible practices.
We test all exploits on a local replication of the Ethereum blockchain that is disconnected from the public network.
Unless the attacks have occurred in the wild and the vulnerabilities are publicly known, we do not disclose the specific addresses of vulnerable contracts in the paper.

\subsection*{Contributions}

To summarize, we make the following contributions:

\begin{itemize}[leftmargin=*]
\item We propose a novel deobfuscation method to remove control flow obfuscation and enable further analysis.
\item We propose a transaction-seeded symbolic execution approach that uses contracts' public historical transactions as high-quality concrete inputs to identify potential vulnerabilities.
\item Building on these two techniques, we implement \name, which takes the bytecode and the historical transactions of a given smart contract as input and outputs exploits if the smart contract is vulnerable\footnote{\name is available at \url{https://github.com/hackingdecentralized/skanf}.}. \name can be used defensively by developers to detect and fix vulnerabilities before deployment or interaction.
\item We evaluate the efficacy of \name against real-world closed-source and obfuscated smart contracts. \name identifies \identifiedbugs vulnerable contracts from \numberofcontracts MEV bots and automatically exploits \confirmedbugs of them. The potential losses of these vulnerabilities exceed \totalloss.
\item We discover 104 real-world attacks targeting real-world contracts, which we categorize as MEV phishing attacks. These have caused an estimated loss of \$2.76M. Of the attacked smart contracts, 27 are flagged by \name.

\end{itemize}

\section{Preliminaries}
\label{sec:background}

\subsection{Ethereum and Smart Contracts}
\parhead{Ethereum}
Ethereum~\cite{wood2014ethereum} is the second-most popular blockchain, and the foremost smart contract-enabled blockchain.
It relies on a Proof-of-Stake (PoS) mechanism, wherein actors can become part of the set of validators operating the blockchain by staking (i.e., depositing) Ethereum's native token, ETH.
Ethereum's mechanism sequentially assigns one validator to propose a block, i.e., a batch of transactions, every 12 seconds.
The validator chosen to act as a proposer has a free hand in constructing its block.

\parhead{MEV supply chain}
Today's Ethereum block construction pipeline involves several specialized entities.
At one end, we have \emph{searchers} who identify profitable opportunities, e.g., transactions that can be front-run~\cite{zhou2023sok}, and assemble transaction bundles that exploit them.
In practice, searchers often automate this process using dedicated smart contracts, which we refer to as \emph{MEV bots}.
These contracts hold assets and implement the logic for interacting with DeFi protocols to execute the identified opportunity on chain.
These bundles are then sent to builders, who specialize in constructing profitable blocks, and who compete in an auction-esque process to have their blocks chosen by upcoming proposers.

\parhead{Smart contract}
Smart contracts are programs stored and executed on the blockchain. They can be invoked through transactions. On Ethereum, contracts are deployed as Ethereum Virtual Machine (EVM) bytecode, typically written in high-level languages such as Solidity~\cite{solidity} and compiled for deployment~\cite{wood2014ethereum}.
Developers may optimize the contracts through compiler settings~\cite{vyper} or by writing bytecode directly.

Executed in the EVM, contracts are sequences of opcodes such as \texttt{CALL} for external interactions and \texttt{JUMP}/\texttt{JUMPI} for unconditional and conditional branching. The EVM is a stack-based environment where operations manipulate a last-in-first-out (LIFO) stack.
Although contracts are stored on-chain as bytecode, their source code and Application Binary Interface (ABI) can be voluntarily published on platforms such as Etherscan~\cite{etherscan2025docs}; without the ABI, it is difficult to determine how to interact with the contract.

\parhead{Indirect jump} 
An \textit{indirect jump} occurs when a jump instruction (\texttt{JUMP}/\texttt{JUMPI}) is executed and the top stack value (i.e., the jump destination) is not a statically known constant, but rather a runtime-computed value.
Obfuscation techniques often rely on indirect jumps to hide control flow, making it difficult to statically determine which path may be executed.

\parhead{External call}
A smart contract can invoke another contract using one of four opcodes: \texttt{CALL}, \texttt{CALLCODE}, \texttt{DELEGATECALL}, and \texttt{STATICCALL}. These opcodes enable the caller to pass along Ether (value), execution gas, and input data (calldata), facilitating inter-contract interactions. They differ in how the execution context is propagated, particularly with respect to \texttt{msg.sender}, \texttt{tx.origin}, and storage access.
In this paper, we focus on the \texttt{CALL} instruction, which underlies common patterns such as calling an ERC-20 token’s \texttt{transfer} function; we refer to such invocations as \textit{external calls}.

The EVM provides two global variables to track the source of a call: \texttt{tx.origin} and \texttt{msg.sender} (accessed via \texttt{ORIGIN} and \texttt{CALLER} opcodes). The variable \texttt{tx.origin} always refers to the original EOA that initiated the transaction, remaining constant across the entire call chain. In contrast, \texttt{msg.sender} refers to the immediate caller of the current function, which may be an EOA or another contract, and it changes with each external call.

For clarity, we refer to \texttt{tx.origin} as \emph{origin} and \texttt{msg.sender} as \emph{caller} throughout the rest of this paper.
We refer to the contract that an external call invokes as the \textit{target address}. The calldata consists of two parts: the first four bytes form the \textit{function selector}, which identifies the \textit{target function} in the target contract; and the remaining bytes, which specify the function arguments.

\parhead{Asset management} In this paper, we focus on ERC-20 token assets held by smart contracts.
The ERC-20 standard~\cite{erc20} supports two transfer mechanisms: an account can call \texttt{transfer} to send tokens directly, or use \texttt{approve} to authorize another account to move tokens with \texttt{transferFrom}.

\subsection{Smart Contract Obfuscation}

Several obfuscation techniques have been proposed, including layout obfuscation~\cite{collberg1997taxonomy,zhang2023bian}, data flow obfuscation~\cite{collberg1997taxonomy,zhang2023bian}, control flow obfuscation~\cite{collberg1997taxonomy,ma2025surviving}, and preventive transformations~\cite{collberg1997taxonomy}.
We focus on tackling control flow obfuscation because it prevents existing detection tools from functioning correctly, and it is commonly present among MEV bots (see~\cite{ma2025surviving} for empirical evidence).

We confirmed that the layout and data flow obfuscation do not interfere with existing analysis tools.
We use the state-of-the-art obfuscator BiAn~\cite{zhang2023bian} to apply layout and data flow obfuscation to 67 vulnerable smart contracts collected from the Smart Contract Weakness Classification registry~\cite{swcregistry}. Mythril can still detect vulnerabilities in all obfuscated contracts.
In the interest of space, we relegate details to~\autoref{sec:obfuscation-tools}.
Preventive transformations, such as self-destructing after execution to make bytecode unavailable, are also less concerning due to blockchain's transparency and immutability; even a self-destructed smart contract's deployment transaction remains permanently recorded and publicly available for analysis.

\subsection{Threat Model}
Our adversary model captures the ability of a typical attacker targeting smart contracts. They can create transactions, deploy smart contracts, and interact with contracts, but cannot access private keys of other accounts, breach the blockchain's integrity, or manipulate the block-building process.

\parhead{Asset management vulnerability}
Smart contracts, when trading on behalf of users, may perform asset management activities.
For example, to execute a token swap across different DEXs, a contract might need to allow external calls from the DEXs, or to implement some pre-defined interface, such as \texttt{uniswapV3SwapCallback}~\cite{IUniswapV3SwapCallback}.
\done%
Our focus is on \emph{asset management vulnerabilities}, i.e., vulnerabilities that would allow an adversary to transfer tokens from the contract's account to the adversary's.
A representative example is discussed in \autoref{sec:destroyer-inu}.
In particular, weak adversaries such as our own can do so when a smart contract relies on obfuscation to hide its lack of access control checks.
Another possibility arises when the smart contract verifies that a token transfer is allowed by checking if the invoking transaction originated from a white-listed address; then, adversaries can employ phishing attacks wherein searchers attempt to interact with a seemingly innocent contract, which then performs the malicious transfer on the adversaries' behalf.

\section{The Design of \name}

\subsection{Overview}
\begin{figure*}
    \centering
    \includegraphics[width=0.9\linewidth]{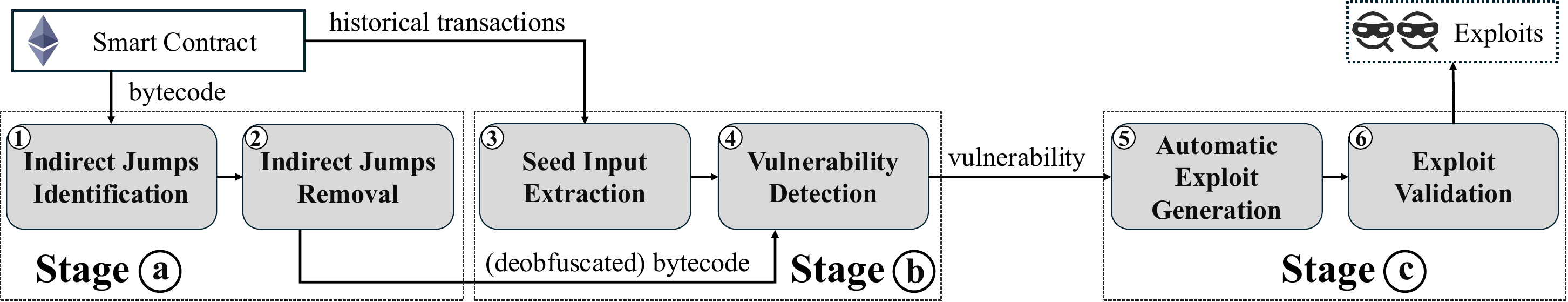}
    \caption{Overview of \name. The input consists of the smart contract's bytecode and historical transactions, while the output is the verified exploit. In the first stage (\textcircled{a}), we identify the obfuscated smart contracts (\ding{192}) and recover their control flow information (\ding{193}).
    In the second stage (\textcircled{b}), \name parses historical transactions as concrete seeds for symbolic execution (\ding{194}). Then it identifies potential vulnerabilities within the contract and labels adversary-controllable parameters (\ding{195}).
    Finally, in the last stage (\textcircled{c}), \name generates potential exploits for the identified vulnerabilities (\ding{196}), which are then validated in \ding{197}.
    }
    \Description{A three-stage pipeline overview of \name. In Stage a, bytecode from a smart contract is processed through indirect-jump identification and indirect-jump removal. In Stage b, historical transactions are used for seed input extraction, followed by vulnerability detection on the deobfuscated bytecode. In Stage c, detected vulnerabilities are passed to automatic exploit generation and exploit validation, producing verified exploits.}
    \label{fig:overview}
\end{figure*}

The inputs to \name are the bytecode of a given smart contract, and, if they exist, historical transactions can also be provided to improve performance.
\done%
\autoref{fig:overview} provides an overview of our system, which consists of the following stages: 

\noindent\textbf{1. Control flow deobfuscation.} First, \name identifies and eliminates control flow obfuscation by reconstructing the \switchtable of an input contract to remove indirect jumps, which allows code analysis tools to operate correctly.

\noindent\textbf{2. Transaction-seeded symbolic execution.} Smart contracts of interest have complex logic and dependencies, requiring symbolic execution guided by historical transactions to speed up vulnerability discovery.
Unlike tools that rely on {\em manual input}~\cite{mythril}, \name automatically identifies high-quality data from historical transactions to guide its exploration of the input contract's possibly large number of logic paths. 
This stage's output is a list of potential vulnerabilities, i.e., insufficiently protected invocations of \texttt{CALL} that could be exploited to steal assets.

\noindent\textbf{3. Exploit generation and validation.} To exploit vulnerabilities identified in the previous stage, \name crafts for each one a transaction that triggers the vulnerable \texttt{CALL} with specific adversarial inputs. A key challenge in this stage is that an exploit should succeed in executing code {\em after} the said vulnerable \texttt{CALL}. To ensure that exploits are valid, \name executes them in a local environment.

\subsection{Control Flow Deobfuscation}
\label{sec:control-flow-deobfuscation}

In \name's first stage, our goal is to deobfuscate the target smart contract, thus facilitating further analysis.
Existing tools generally fail to analyze obfuscated smart contracts, making deobfuscation a prerequisite for downstream vulnerability analysis.
Technically, we need to identify indirect jumps (i.e., whose destinations are determined at runtime) and convert them to \emph{direct} ones (whose destinations are statically specified) to make the control flow explicit and statically analyzable.
Note that even with transaction-seeded symbolic execution, this step remains necessary. Because the seeds only initialize inputs and execution stays symbolic, indirect jumps can still hinder exploration; deobfuscation resolves these jumps and enables continuous exploration toward the target logic.

\begin{figure}[thbp]
    \centering
    \includegraphics[width=0.9\linewidth]{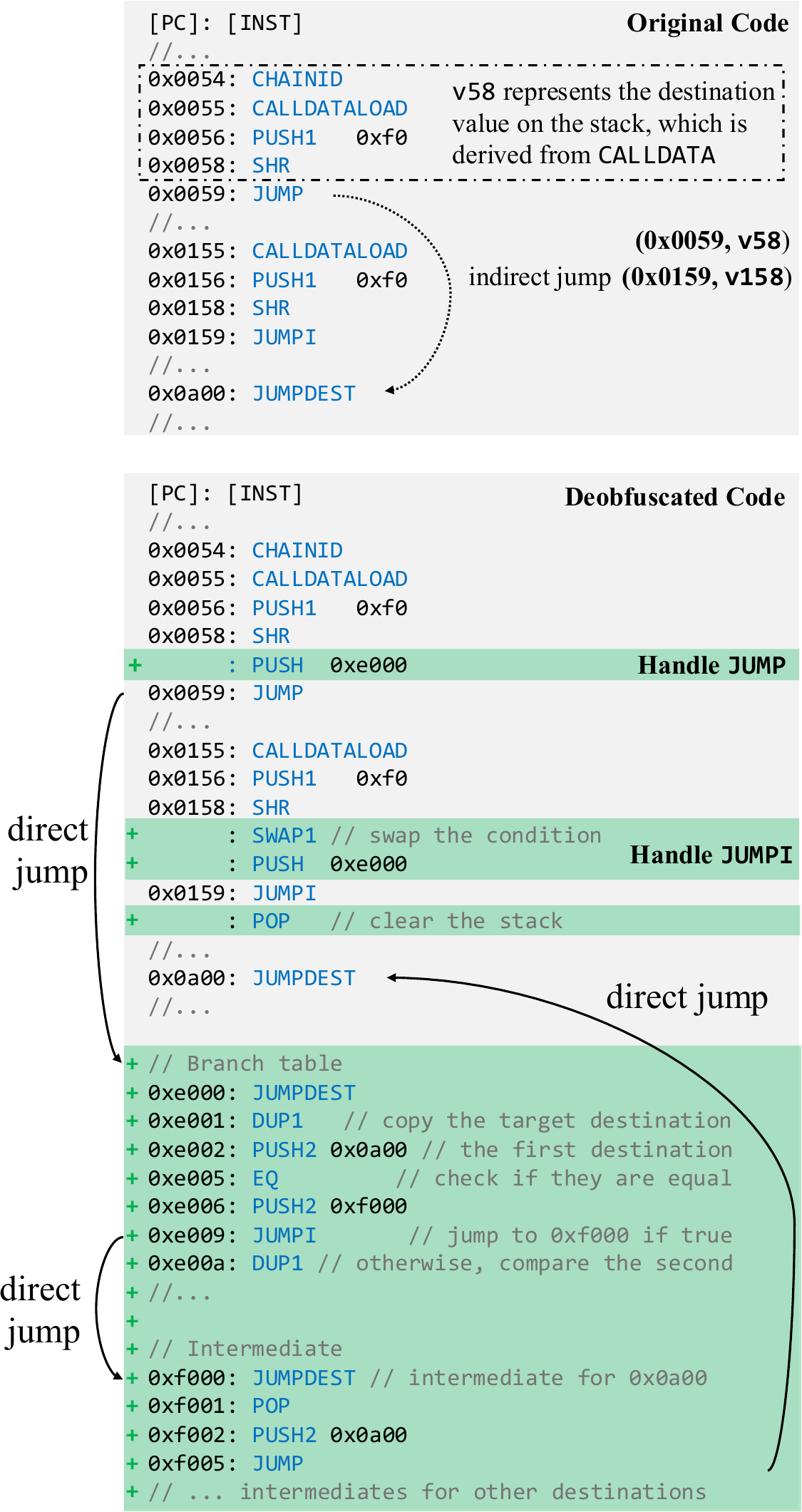}
\caption{Assembly code for the obfuscated smart contract and how to rewrite an indirect jump as a direct one by inserting a crafted \switchtable. The added lines do not carry a PC value, nor do they affect the PC of existing ones (see \autoref{sec:control-flow-deobfuscation}).
\done%
}
\Description{A two-part assembly illustration of control-flow rewriting. The top part, labeled Original Code, shows EVM instructions in which calldata is loaded, shifted, and then used by JUMP and JUMPI as indirect destinations. The bottom part, labeled Deobfuscated Code, shows added instructions highlighted in green. These inserted instructions route execution to a new branch table and intermediate blocks, where the destination is checked and then translated into direct jumps. Curved arrows indicate the correspondence between the original indirect jumps and the new direct-jump structure.}
\label{fig:assembly-of-switch-table}
\end{figure}

\parhead{Identifying indirect jumps}
We begin by scouring the input bytecode for indirect jumps. A notable difficulty that is faced when ``cleaning up'' obfuscated control flow logic is that the destination of an indirect jump may be based on dynamic inputs such as \texttt{calldata}.
\done%
To identify such jumps, we extract all jump instructions (\texttt{JUMP} and \texttt{JUMPI}) and build a control flow graph (CFG).
We then analyze whether each jump's destination depends on dynamic inputs (i.e., \texttt{calldata}, \texttt{value}, \texttt{memory}, and \texttt{storage}).
Indirect jumps are recorded in a set $\mathcal{J} = \{(j_1, d_1), \dots, (j_n, d_n)\}$,
where $j_i$ is the PC of the jump instruction and $d_i$ is the stack variable that determines the jump destination at runtime (e.g., \texttt{v58}, \texttt{v158} in~\autoref{fig:assembly-of-switch-table}).

If $\mathcal{J}$ is empty, the contract does not use control flow obfuscation, and \name continues to the second stage.
Otherwise, we proceed to deobfuscate the indirect jumps.

\parhead{Deobfuscation}
The EVM's specification requires that every valid jump destination must be explicitly marked with a \texttt{JUMPDEST} instruction~\cite{wood2014ethereum}. %
This property allows us to statically scan the disassembled bytecode, collect all legal jump destinations, and instrument the bytecode with a \switchtable that replaces indirect jumps with direct jumps to restore analyzable control flow.
This transformation preserves the original control flow while ensuring that jump destinations are constants encoded in the bytecode.

The idea is best illustrated with an example. 
Suppose the original contract contains an indirect jump to an address stored in variable \texttt{v}. 
Most static analysis tools struggle to handle this since there are too many possible branches.
Our idea is to rewrite the code as follows:
\begin{lstlisting}[style=YulStyle,xleftmargin=5.0ex]
if v == 0x0a00 : jump 0x0a00
else if v == 0x0b00 : jump 0x0b00
else if v == ...
\end{lstlisting}
where \texttt{0x0a00}, \texttt{0x0b00}, and so on, are jump destinations extracted from the bytecode. 
The number of branches is exactly the number of valid jump destinations.

\done%

Replacing every jump instruction with a copy of the above code is inefficient, so we reuse the code by adding a {\em \switchtable}.
A \switchtable is a structure commonly used to implement multi-way branching based on a runtime value. 
We implement the table as a sequence of conditional checks on the variable holding jump destinations (e.g., \texttt{v58} in~\autoref{fig:assembly-of-switch-table}).
For each valid destination address, we populate the table with an entry that jumps to the address if the variable is equal to it; otherwise, it proceeds to the next entry.
In doing so, we take the contract's ``implicit'' and opaque control flow where destinations are ``tucked away'' in a variable, and make it \emph{explicit}: all possible valid destinations are now clearly enumerated in the text.
To ensure that the added code does not interfere with existing logic, we cleverly rely on the 24KB (0x6000 bytes) size limit of contract bytecode~\cite{buterin2016eip170}, and insert our tables at PC \texttt{0xe000}.
Note that any position after \texttt{0x6000} could work.

\done%

If the original instruction is \texttt{JUMPI}, special handling is required because it involves two stack values: the jump destination (on top of the stack) and the condition (just below it).
To preserve the original semantics when redirecting execution to the \switchtable, we must maintain the correct stack layout.
Before inserting the new jump destination, we insert a \texttt{SWAP1} to move the condition to the top of the stack.
This ensures that when the rewritten \texttt{JUMPI} executes, it evaluates the condition correctly while using the \switchtable entry as the destination.
We also insert a \texttt{POP} immediately after \texttt{JUMPI} to remove the unused destination value in case the condition is false and the jump does not occur, thereby avoiding stack imbalance.

To appreciate the subtleties in seamlessly replacing indirect jumps, we need to dive into the technicalities involved.
For illustration, we refer at each step to the corresponding lines in a corresponding example given in~\autoref{fig:assembly-of-switch-table}.
For every indirect jump instruction at PC $j \in \mathcal{J}$, we ``redirect'' the jump destination to our table by rewriting the bytecode to push \texttt{0xe000} onto the stack.
Each table entry is implemented to account for the stack-based semantics of the EVM:
First, the runtime value is duplicated (see PC \texttt{0xe001}), and a known destination is pushed (e.g., \texttt{0x0a00} in~\autoref{fig:assembly-of-switch-table}), allowing us to compare the two using the \texttt{EQ} instruction (PC \texttt{0xe005}).
If both are equal, the program jumps to an intermediate ``gadget'' which cleans up the stack and performs the final jump (see PC \texttt{0xf000}–\texttt{0xf005}).
This intermediate step is necessary because the comparison only consumes the duplicated value, leaving the original runtime value on the stack.
If the comparison fails, execution continues to the next table entry (e.g., at PC \texttt{0xe00a}), which repeats the process for another possible destination.
This continues until a matching entry is found.

\done%
In order to prevent bytecode offset shifts, we do not directly modify the raw bytecode when instructions need to be injected at PC values lower than \texttt{0xe000} (e.g., consider the \texttt{PUSH} instruction between PC values \texttt{0x0058} and \texttt{0x0059}).
Instead, we apply such changes only when the analysis of the program operates on a CFG rather than concrete bytecode offsets.
As a result, injected instructions do not carry PC values.
In addition, the \switchtable includes all possible jump destinations within the contract.
Although some destinations may not be used at runtime, the evaluation in~\autoref{sec:evaluation-deobfuscation} shows that this does not introduce significant overhead.

After this transformation, all jump targets in the contract are encoded as immediate values, and the control flow becomes fully explicit and statically analyzable.

\subsection{Vulnerability Detection}
\label{sec:vulnerability-detection}
In this stage, we identify potential asset management vulnerabilities in the smart contract---specifically, opportunities for an adversary to steal assets.
We define a \emph{vulnerability oracle} that flags any \textit{vulnerable \texttt{CALL} instruction} as one that is (1) reachable from a public entry point and (2) accepts inputs that are either adversary-controlled or fixed but risky (e.g., a known ERC-20 token address).

Note that a vulnerability does not guarantee a successful exploit.
This is akin to traditional vulnerabilities like buffer overflows, where a flaw alone does not guarantee arbitrary code execution, as additional conditions may be required.
In this paper, we define an exploit as a transaction that transfers assets from the victim contract to the adversary's account.

We detect vulnerabilities in this stage and generate exploits in the next.
To this end, we first extract concrete seed inputs from historical transactions to perform seeded symbolic execution, thereby speeding up the exploration.
Additionally, we perform taint analysis during the path exploration leading to a vulnerable \texttt{CALL} instruction.
The taint analysis results are then used to determine whether the reachable \texttt{CALL} instruction is one of the vulnerabilities of interest.

\parhead{Extracting seed input from historical transactions}
A key challenge in analyzing real-world smart contracts lies in their complex logic, especially when involving multiple \texttt{CALL} instructions to interact with other smart contracts, such as DEXs, NFTs, and lending protocols.
To efficiently explore such code, we adopt \textit{seeded symbolic execution}~\cite{sen2007concolic}, which initializes symbolic execution with concrete \textit{seed inputs} to guide path exploration and simplify constraint solving, thereby reducing the search space and improving efficiency. The idea of seeded execution itself is not new; our key observation is that, for closed-source smart contracts, public historical transactions provide an effective source of high-quality seeds. We do not replay full concrete executions; the seeds only initialize the selected inputs, while the exploration remains symbolic.

In our setting, an ideal seed input is a set of parameters that trigger an external call to transfer assets out of the smart contract.
This prioritizes the exploration of ``high-risk'' paths involving asset transfers.
Crafting seed inputs for closed-source smart contracts is hard because their ABI is unknown.
Randomly generated seed inputs lack insight into the code's logic, and thus may have a low probability of reaching vulnerable paths.
\done%
Fortunately, high-quality seed inputs are already readily available on-chain in historical transactions.
Given a target contract, we retrieve the associated transactions that are executed successfully and extract the relevant inputs, which serve as concrete seeds for seeded symbolic execution.

Focusing on asset-transferring paths may introduce false negatives, since not all potentially vulnerable paths are explored.
However, this trade-off is acceptable---discovering even a single asset management vulnerability suffices to demonstrate that the contract's assets are at risk.
As shown in~\autoref{sec:vulnerability-detection-evaluation}, this approach allows us to uncover more vulnerabilities than existing tools.

Specifically, we collect historical transactions of a given contract $C$, re-execute them on the corresponding historical state using an archive node~\cite{ethereum2025archive}, and extract external calls that involve ERC-20 transfers by inspecting standardized \textit{transfer event logs}~\cite{erc20}.
For relevant transactions, we extract all associated \contractcall{}s, keeping those targeting $C$.
We further filter out calls with empty calldata (e.g., plain ETH transfers), which typically do not involve asset management logic.

\parhead{Seeded symbolic execution}
From the parsed historical transactions, we extract all \contractcall{}s to the target contract, including the origin, caller, calldata, and value. These, along with the block number which reflects the correct blockchain state, are used as concrete inputs for seeded symbolic execution.

While seeded symbolic execution can help identify reachable \texttt{CALL} instructions, further analysis is required to determine whether the inputs to these instructions are vulnerable. 
During execution, we apply \textit{on-the-fly taint propagation} with \textit{byte-level} granularity to track how tainted calldata bytes, representing adversarial inputs, flow into critical inputs of \texttt{CALL} instructions.
Specifically, we decompose the input to the \texttt{CALL} instruction into different \textit{parameters}, including the target address, the function selector, and individual function arguments.

Taint propagation is initiated when calldata is accessed (e.g., via \texttt{CALLDATALOAD} or \texttt{CALLDATACOPY}) and proceeds dynamically through the EVM's stack, memory, and storage.
Our policy is conservative: taint is propagated through all arithmetic, logical, and data copy instructions.
For each instruction, if any operand is tainted, the resulting value is also marked as tainted. %
We treat external calls as taint sinks and examine if the target address, function selector, or any arguments are influenced by tainted data.
Further details of our taint tracking are provided in~\autoref{sec:taint-tracking}.

When seeded symbolic execution encounters a \texttt{CALL}, it performs two key operations. First, we inspect the instruction's parameters to determine whether it represents a potential asset management vulnerability. Then, based on taint analysis, we selectively replace calldata bytes that influence the \texttt{CALL} parameters with symbolic variables, while preserving concrete values for all other bytes. In stage \textcircled{c}, we generate exploits by restarting seeded symbolic execution with these modified calldata bytes, which allows us to construct an exploit based on the values specified by the adversary.

\parhead{Identifying asset management vulnerability}
After the taint analysis, \name identifies \texttt{CALL} parameters that can be determined by the adversary via their input. We refer to such parameters as \textit{adversary-controllable}.
However, even when some parameters are fixed (thus cannot be changed by an adversary), the \texttt{CALL} instruction may still be vulnerable. For example, the function selector (0xa9059cbb, corresponding to \texttt{transfer}) is fixed, yet the adversary still controls the target address of the call and can construct the calldata to transfer all WETH to their account. Therefore, we also analyze fixed parameters to assess whether they pose a risk.

In the context of asset management vulnerabilities, a fixed target address may still be problematic if it corresponds to an ERC-20 token contract (e.g., WETH).
Similarly, a fixed function selector may be risky if it maps to functions like \texttt{transfer} or \texttt{approve}.
We call such parameters \textit{risky}, as they can lead to asset loss even without full adversarial control.

Once we determine that both the target address and function selector of the vulnerable \texttt{CALL} instruction are either adversary-controllable or risky, we examine if bytes 5 to 36 of its calldata, which are interpreted as the first argument if the function is \textit{transfer} or \textit{approve}, can be arbitrarily set by the adversary, noting that if the function selector is controllable, the adversary can choose \textit{transfer} or \textit{approve}.
If this is the case, we classify the corresponding \texttt{CALL} instruction as a potential asset management vulnerability. Additionally, we record whether the second argument (i.e., the transfer amount) is fixed or under adversarial control.

\parhead{Fallback mode}
Given that not all contracts are highly active and may lack sufficient historical transactions, we incorporate symbolic execution as a fallback when seeded symbolic execution fails to identify potential vulnerabilities. 
In fallback mode, \name conducts a broader analysis to identify all potentially vulnerable \texttt{CALL} instructions and assess their reachability.
Specifically, we first enumerate all \texttt{CALL} instructions in the contract.
For each instruction, we use the call graph to identify functions that are reachable from public entry points, and the CFG to determine whether those functions can eventually reach the \texttt{CALL} instruction. 
If such a path is found, we initiate symbolic execution from the corresponding public function using a symbolic byte string as calldata.
\done%

As symbolic execution progresses, we keep track of symbolic expressions for all values in the EVM stack, memory, and storage.
These symbolic values are updated with each instruction to reflect how they depend on input variables.
When the execution encounters a conditional branch (i.e., \texttt{JUMPI}), we fork the execution into two paths---one for each possible outcome---and add the corresponding condition to the path constraints.
To manage complexity, we prune infeasible paths.
This can happen in two cases: either the accumulated constraints become unsatisfiable (i.e., no input could result in that path), or the CFG shows that the path cannot reach the target \texttt{CALL} instruction. 
Handling access control checks---like those involving \textit{msg.sender} or \textit{tx.origin}---can be particularly tricky.
To improve the chance of passing these checks, we employ two configurations.
First, we set both the caller and the origin to a predefined adversary address.
If that fails, for example, when the contract requires a specific \textit{tx.origin}, we perform data-flow analysis on the value compared against the \texttt{ORIGIN} opcode and extract the corresponding address from the contract for a second attempt.

Once symbolic execution discovers a feasible path to a \texttt{CALL} instruction, we apply the same criteria as in seeded symbolic execution to determine whether any parameters are adversary-controllable or otherwise risky.

\done%
\parhead{Preliminary validation}
For each vulnerability, we perform a preliminary validation step before attempting to exploit it, as concrete exploit generation requires additional constraints to be satisfied.
This validation is worthwhile because symbolic execution may incorrectly identify vulnerabilities even when these cannot be exploited, due to imprecise modeling of external calls and incomplete representations of storage or execution context~\cite{gritti2023confusum}.
For instance, a \texttt{CALL} instruction may appear reachable in symbolic analysis but cannot be triggered under any concrete execution, as symbolic execution may over-approximate feasible paths when constraints are incomplete.

Our preliminary validation uses a local blockchain instance to simulate a concrete execution environment.
We supply concrete calldata, solved from symbolic constraints, along with caller and origin addresses to verify whether the \texttt{CALL} instruction is executed.
Unlike exploit generation, we do not require the transaction to succeed or transfer assets; we only check whether the vulnerable \texttt{CALL} is triggered.
This suffices to confirm the vulnerability's existence.

At the end of this stage, \name produces a ``vulnerability report'' (an example is given in~\autoref{fig:vulnerability-output}), recording the PC of the \texttt{CALL} instruction, the risky or adversary-controllable parameters, and other contextual information such as the caller, origin, block height, and the concrete calldata used.
This is then forwarded to the next stage for exploit generation.

\begin{figure}
    \centering
\begin{lstlisting}[language=json]
{
  "caller": "0xdead...beef", 
  "origin": "0xdead...beef",
  "blockNumber": 20000000,
  "callPC": "0xac5", // PC of the vulnerable CALL
  "calldata": "12345678SS...", // SS is the symbolic data
  "targetAddress": "*", // adversary-controllable
  "functionSelector": "0xa9059cbb", // fixed but risky 
  "destination": "*",
  "amount": "*"
}
\end{lstlisting}
    \caption{Example output of \name for a vulnerability.}
    \Description{A JSON-like example output produced by \name for a detected vulnerability. The output includes fields for caller, origin, block number, program counter of the vulnerable CALL, calldata, target address, function selector, destination, and amount. Some fields are concrete values, while others are marked with wildcards to indicate symbolic or adversary-controllable inputs. Comments note that the program counter identifies the vulnerable CALL, the calldata contains symbolic data, the target address is adversary controllable, and the function selector is fixed but risky.}
    \label{fig:vulnerability-output}
\end{figure}

\parhead{Mitigating state explosion}
\revision{
The state explosion originates from the indirect jumps in obfuscated contracts, as the jump targets are computed at runtime and need to be explored symbolically.
Our introduction of a \switchtable does not create this issue; rather, it makes the possible jump targets explicit and thus analyzable.
For seeded symbolic execution, this concern is mitigated because exploration is guided by concrete traces from historical transactions. For pure symbolic execution, however, we may need better strategies.
}

To maintain analytical scalability, we implement a path pruning heuristic that caps the number of visits to the \switchtable within a single execution path.
Because all indirect jumps pass through this table, limiting repeated entries is both effective and safe: such indirect jumps are used to enter self-contained logic blocks, so repeatedly jumping to the same block within a single trace may not contribute new behavior. Bounding such revisits therefore reduces cyclic exploration while preserving the contract's core logic.
As discussed in~\autoref{sec:pruning-rule}, our evaluation shows that this heuristic improves scalability in practice without affecting vulnerability detection.

\done%

\subsection{Exploit Generation and Validation}
At the final stage, \name attempts to synthesize and validate a concrete exploit for each identified vulnerability.

\parhead{Automatic exploit generation}
Given a vulnerability identified in the previous stage, \name attempts to construct a transaction that successfully exploits it.
The transaction is built at the specific block height where the vulnerability is observed.
The ``to'' address is set to the victim contract.
The ``gas'' field is set using the node's estimate for successful execution at that block height, and the fee parameters are set based on the block's base fee so that execution cost is not the limiting factor in our validation.
The ``value'' field is treated symbolically and solved during exploit generation; when no positive value is required, it is set to zero.
The ``from'' address is set to either the adversary's address (if no constraints are specified) or a specific origin address if one was recorded during vulnerability detection.
The latter case is typically caused by tx.origin-based authorization, which can be bypassed through phishing attacks, as is the case with the Destroyer Inu attack presented in \autoref{sec:destroyer-inu}.
Given how easy such phishing attacks are in practice (see~\autoref{sec:evaluation-attacks} for real-world evidence), we conservatively assume that the attacker can always set the ``from'' value to the required one.

The key step in exploit generation is constructing the correct calldata to ensure the transaction executes successfully. 
Recall from stage \textcircled{b} that for each vulnerability, we record which \texttt{CALL} parameters---the target address, function selector, and arguments---are adversary-controllable.
If the target address is controllable, we constrain it to the address of an ERC-20 token held by the victim contract with a non-zero balance. By repeating the attack with different addresses, we can systematically target each token.
Similarly, if the function selector and arguments are controllable, we constrain them to values that mimic a realistic attack: we set the selector to \texttt{transfer}, the recipient (\texttt{to}) to the adversary's address.
For the transfer amount, we query the token balance of the victim contract at the specified block height and set this balance as the transfer amount, aiming to steal the maximum possible amount as an adversary would.

According to the EVM specification~\cite{wood2014ethereum}, a transaction is considered successful only if execution halts at a valid stop instruction (\texttt{STOP} or \texttt{RETURN}).
To ensure this condition is met, we resume seeded symbolic execution from the vulnerable \texttt{CALL} and symbolically execute forward until a stop instruction.
This introduces additional path constraints, which are solved to adjust the synthesized calldata and complete exploit generation.

\parhead{Exploit validation}
Finally, \name validates the exploit in a local execution environment to ensure that it executes successfully and produces the expected behavior.

To do this, we simulate the blockchain environment at the block height specified in the exploit.
\revision{The precise historical state is retrieved from an archive node, which allows us to load the exact contract code, account balances, and persistent storage state.}
We then execute the synthesized transaction.
Finally, we check whether the transaction succeeded, and further assess exploit success by inspecting the emitted event logs: for ERC-20 exploits, we verify the presence of a \texttt{Transfer} or \texttt{Approval} event consistent with the expected asset movement, following the ERC-20 standard~\cite{erc20}. If both checks pass, \name confirms the exploit as valid.

\section{Evaluation}
\label{sec:evaluation}

\subsection{Experimental Setup}
\label{sec:eval-setup}

\parhead{Implementation}
We implement \name on top of Gigahorse~\cite{grech2019gigahorse,grech2022elipmoc} and Greed~\cite{gritti2023confusum,ruaro2024not,ruaro2025history,ruaroapprove}, comprising 2.2K lines of Python code and 210 lines of Soufflé code.
In more technical detail, we develop a Gigahorse plugin with two main tasks. First, it analyzes bytecode to identify indirect jumps that indicate control-flow obfuscation. Second, for contracts identified as obfuscated, it recovers the control flow as described in~\autoref{sec:control-flow-deobfuscation}.
We then use Gigahorse to lift the (deobfuscated) bytecode to its register-based Intermediate Representation (IR).
On top of this IR, we build on Greed, a symbolic execution engine, and implement components for seeded symbolic execution, taint analysis, and automatic exploit synthesis to support vulnerability detection and exploit generation.
The system attempts to synthesize exploits through constraint-solving given the following inputs: the register-based IR of a contract along with the CFG and call graph generated by Gigahorse.
For a given contract, we fetch its recent transactions using Etherscan's API~\cite{etherscan2025docs}.
To validate exploits, we reconstruct the historical state using a Reth archive node~\cite{reth} and py-evm~\cite{pyevm,ethpwn}, and simulate their execution.

\parhead{Dataset}
We use real-world MEV bots as the target to evaluate the efficacy and performance of \name. 
We compile a dataset of \numberofcontracts MEV bot addresses 
from multiple sources, including prior works~\cite{yang2024decentralization,oz2024wins,heimbach2024nonatomic}, public dashboards~\cite{LibMEVLeaderboard2025,etherscan2025mevbot}, and results from MEV analysis tools such as MEV inspect~\cite{flashbots2024mevinspect}, within which we identified 5,389 (82.2\%) contracts with unique \texttt{SHA256} hashes of bytecode.

\parhead{Testbed}
Our experiments are conducted on a server running Ubuntu 22.04, equipped with an Intel Xeon Platinum 8380 processor (80 cores), 128GB of RAM, and an 8TB SSD.

\subsection{RQ1: Deobfuscation Effectiveness}
\label{sec:evaluation-deobfuscation}

The effectiveness of \name in addressing control-flow obfuscation is measured using the \textit{code coverage} metric.
Here, \textit{code coverage} refers to CFG completeness in static analysis, that is, the fraction of basic blocks that are statically reachable in the recovered control-flow graph, rather than runtime execution coverage.
Accordingly, a tool's code coverage for a given smart contract is defined as the fraction of statically reachable code blocks in the CFG recovered by that tool.
Low code coverage indicates that a significant portion of the smart contract cannot be analyzed by that tool.

We compare Gigahorse's coverage with and without deobfuscation for all smart contracts in our dataset.
To zoom in on real-world obfuscation practices used by major actors, we repeat the comparison on the top 10 and top 100 most active MEV bots since the Merge (ranked by the number of bundles they sent), as reported by the ``libMEV'' online dashboard~\cite{LibMEVLeaderboard2025}.
We choose these bots because sophisticated MEV searchers who remain active are strongly motivated to keep their business logic private, making them more likely to adopt control flow obfuscation in their code.

\begin{figure}
    \centering
    \includegraphics[width=\linewidth]{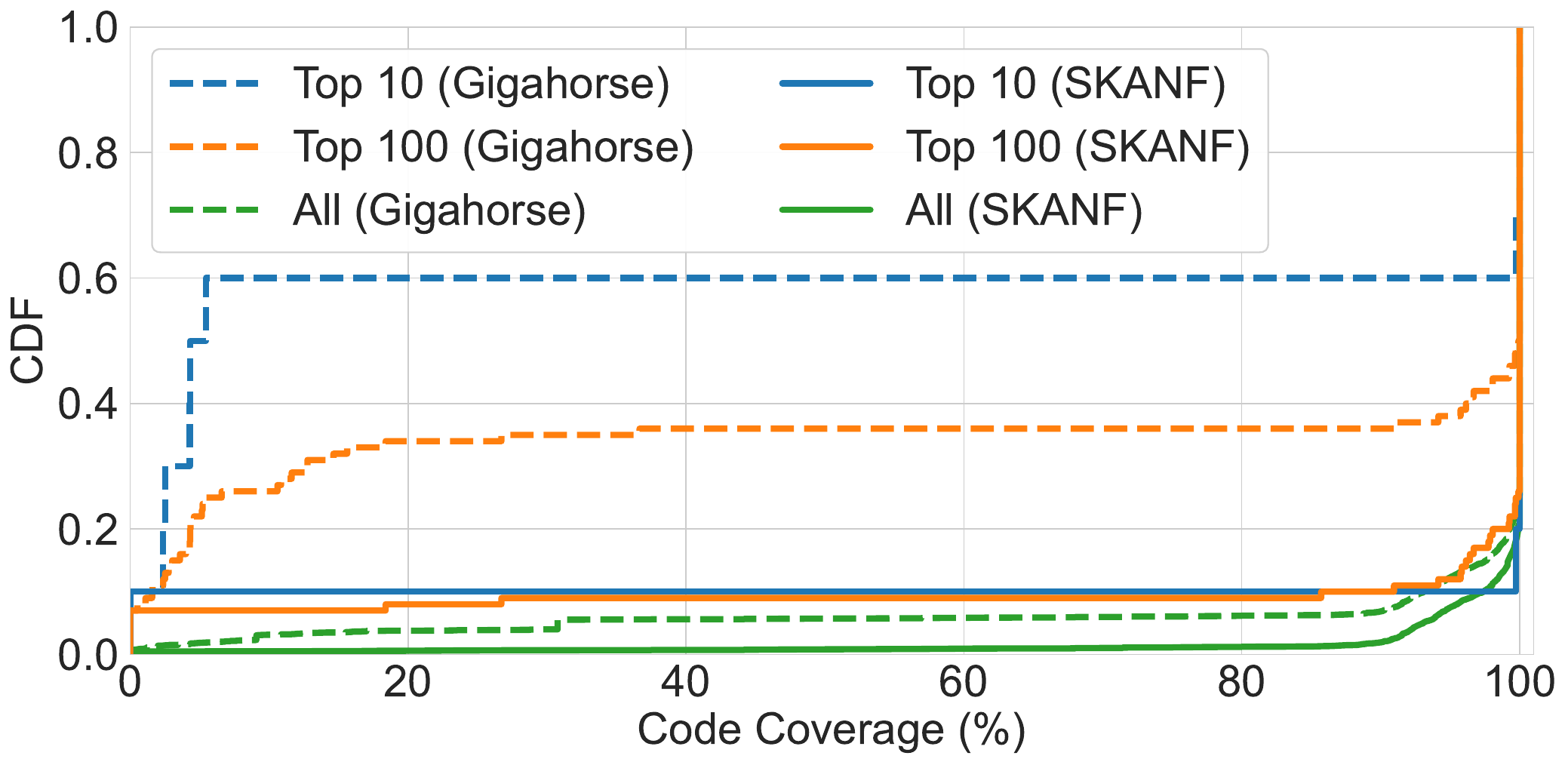}
    \caption{Cumulative Distribution Function (CDF) of the code coverage of smart contracts in our dataset for both Gigahorse and our tool, \name. As can be observed, \name's CDF is concentrated near 100\%, while that of Gigahorse is spread out and is consistently higher (i.e., \emph{worse}) than \name, indicating that \name outperforms Gigahorse.}
    \Description{A cumulative distribution plot comparing code coverage between Gigahorse and \name. The x-axis shows code coverage from 0 to 100 percent, and the y-axis shows the cumulative fraction of contracts. Six curves are shown: Top 10, Top 100, and All for each tool. The \name curves remain close to full coverage for most contracts, whereas the Gigahorse curves are distributed over a wider range of lower coverage values.}
    \label{fig:code-coverage}
\end{figure}

\autoref{fig:code-coverage} presents the cumulative distribution function (CDF) of code coverage for the top 10, top 100, and all smart contracts in the dataset, respectively. Dashed lines represent Gigahorse, and solid lines represent \name.
Among the top 10 MEV bots, six are highly obfuscated (with code coverage below 10\%), and \name successfully increases the code coverage to 100\% for five of them.
The performance on the remaining bot contract is due to the reliance of \name on preliminary output from Gigahorse for bytecode analysis. Although Gigahorse generally produces preliminary output even for highly obfuscated contracts, in this case, it does not do so, which prevents \name from proceeding.
Comparing the code coverage of three MEV bot groups (top 10, top 100, and all bots), we find that control flow obfuscation is more prevalent among the most active MEV bots, underscoring our approach's importance.

For all bots, we observe that for 5\% of the smart contracts, the original code coverage is below 50\%. In contrast, fewer than 0.4\% of contracts have deobfuscated code coverage below 50\%. This highlights the effectiveness of \name in mitigating obfuscation.
For the small fraction (0.4\%) of contracts where deobfuscation does not significantly improve coverage, our manual analysis reveals two reasons: either Gigahorse fails to produce any preliminary output before deobfuscation, or some blocks within the contracts appear to be unreachable, rather than hidden by control flow obfuscation.

\begin{tcolorbox}[title=Answer to \textbf{RQ-1}, left=2pt, right=2pt, top=2pt,bottom=2pt]
\name effectively addresses control flow obfuscation in smart contracts; for 90\% of contracts with initial coverage below 50\%, we can successfully increase code coverage.
\end{tcolorbox}

During our evaluation, we also counted the number of branches in each \switchtable (see~\autoref{sec:branch-table-study} for details). Overall, among all obfuscated smart contracts in our dataset, about 38\% have fewer than 100 branches, and the median number of branches is 172.
This indicates that the complexity of the inserted \switchtable is modest, and none caused a timeout during our evaluation.

\subsection{RQ2: Vulnerability Detection Effectiveness}
\label{sec:vulnerability-detection-evaluation}
We now evaluate the effectiveness of \name in detecting vulnerabilities in smart contracts, as measured by the number of identified vulnerabilities.
To the best of our knowledge, no existing tool can detect vulnerabilities in control-flow-obfuscated smart contracts. To benchmark our tool, we conduct a comparative evaluation against three of the best available state-of-the-art bytecode-based symbolic execution tools: Mythril (commit: 9e9ee39)\cite{mythril}, ETHBMC (commit: e887f33)\cite{frank2020ethbmc}, and JACKAL (commit: 3993e5c)~\cite{gritti2023confusum}. These tools provide different levels of support for asset-management vulnerability detection and exploit construction. ETHBMC and JACKAL support automated exploit synthesis, while Mythril reports concrete triggering inputs, such as calldata, but does not synthesize a full exploit. These concrete inputs make Mythril useful for assessing whether a reported issue can be triggered in execution. We also include Securify (commit: 51ba124)~\cite{tsankov2018securify} as a static analysis baseline, as it covers bytecode-level security patterns relevant to the vulnerabilities considered in our evaluation.
In the comparative evaluation, we run each tool for a maximum of 10 minutes per contract and impose a memory limit of 20 GB.

\subsubsection{Vulnerability detection by \name}
To illustrate how seeded symbolic execution enhances the effectiveness of vulnerability detection, in our evaluation, we use the fallback mode in \autoref{sec:vulnerability-detection}---i.e., full symbolic execution without concrete inputs---as the \textit{baseline mode} for ablation.
In our evaluation, \name detects 738 vulnerable smart contracts in the baseline mode.
When \name is executed in seeded mode, the number of detected vulnerable smart contracts increases to \identifiedbugs.
For 54\% of the unique vulnerable contracts identified by seeded symbolic execution, \name timed out in the baseline mode. In the remaining cases where symbolic execution fails, we find that the corresponding path constraints are incorrectly deemed unsatisfiable. In contrast, seeded symbolic execution, guided by concrete inputs, can avoid this limitation and successfully explore the vulnerable paths.
This highlights a significant improvement in vulnerability detection achieved through seeded symbolic execution, which simplifies constraint solving and prioritizes exploration of the vulnerable \texttt{CALL} instruction within the limited time.

Since the compared tools do not support deobfuscation, we also report results on the subset of 5,784 contracts that do not require deobfuscation. This comparison isolates the effect of seeded symbolic execution, since control-flow reconstruction is not needed for this subset. On these contracts, the baseline mode detects 537 vulnerable contracts, whereas the seeded mode increases the number to 801. Thus, this improvement can be attributed to seeded symbolic execution, while the remaining gains on obfuscated contracts reflect the additional benefit of control-flow reconstruction.

\parhead{Comparison with state-of-the-art tools} 

{\bf Mythril} alerts when it detects a publicly reachable \texttt{CALL} instruction where the adversary can control the target address.
In the evaluation, Mythril flags 576 potentially vulnerable smart contracts.
We validate these identified vulnerabilities by executing the exploit transaction it generates in a local blockchain instance, similar to our preliminary validation for \name.
Specifically, we check whether the vulnerable \texttt{CALL} instruction is triggered by the adversary and whether the adversary can control other parameters to steal the contract's assets.

In the end, we only reach the \texttt{CALL} instructions in 83 of the flagged contracts (\name identifies \identifiedbugs). Among these, 64 are also covered by \name. For the rest not identified by \name, we manually inspect them and find that identifying vulnerabilities requires an accurate CFG. However, we construct the CFG using Gigahorse, which fails on these contracts (e.g., due to out-of-memory errors during bytecode lifting).

{\bf ETHBMC} also raises an alert when it detects a \texttt{CALL} where the adversary controls the target address and transfer amount~\cite{frank2020ethbmc,gritti2023confusum}.
Note that this captures only a subset of asset management vulnerabilities, as an adversary can still steal assets even if the target and amount are fixed, as long as they can control the recipient.

However, in our evaluation, ETHBMC did not detect {\em any} vulnerabilities.
Our further analysis shows that low code coverage and timeouts are the two main reasons.
For about 36\% of the contracts, ETHBMC achieved less than 25\% code coverage during execution. For about 27\% of the contracts, ETHBMC encountered timeout errors while solving constraints.
These findings highlight ETHBMC's limitations in analyzing obfuscated smart contracts and identifying asset management vulnerabilities in contracts with complex logic.

{\bf JACKAL} detects reachable \texttt{CALL} instructions where the adversary controls both the target address and the function selector.
For the same reason as ETHBMC, this pattern represents a subset of asset management vulnerabilities. In our evaluation, JACKAL detects 18 vulnerable contracts, all within the \identifiedbugs contracts we identified.

{\bf Securify} includes a pattern for \textit{unrestricted ether flow}~\cite{tsankov2018securify}, which is relevant to our setting because it flags potentially unsafe ether-transfer behavior.
Note that this pattern is limited to unsafe ETH flow, whereas \name targets a broader class of asset management vulnerabilities, including those involving ERC-20 token transfers.

In our evaluation, Securify flagged 123 contracts in our dataset for this pattern, all of which belong to the unobfuscated subset of 5,784 contracts. Among them, 45 were also reported by \name. Another 48 were initially reported by \name but excluded during preliminary validation because the corresponding \texttt{CALL} instructions were unreachable. For the remaining contracts, false positives could not be directly ruled out because Securify only reports pattern matches and does not provide candidate calldata for validation.
As noted earlier, \name identified 801 vulnerable contracts on the same unobfuscated subset, showing that its gains over baseline tools are not solely due to control-flow deobfuscation.

To further assess the remaining cases, we performed additional case-by-case analysis using the decompiled code from Dedaub~\cite{dedaub} to determine whether these alerts were likely false positives.
Among them, 8 could not be analyzed further because Dedaub did not provide decompiled code. Of the analyzable contracts, 21 were confirmed as false positives through manual analysis, as the corresponding \texttt{CALL} instructions were guarded by restrictions on \texttt{msg.sender}.

The last case was a true positive missed by \name.
In this contract, the recipient of the transfer is derived from the return value of an external contract call, where the callee address is controllable through calldata and can therefore be malicious.
This reveals a current limitation of \name: it does not consider cases where adversarial control is mediated indirectly through the return value of an attacker-influenced external call.

\subsubsection{Runtime of \name} 
\begin{figure}
    \centering
    \includegraphics[width=0.95\linewidth]{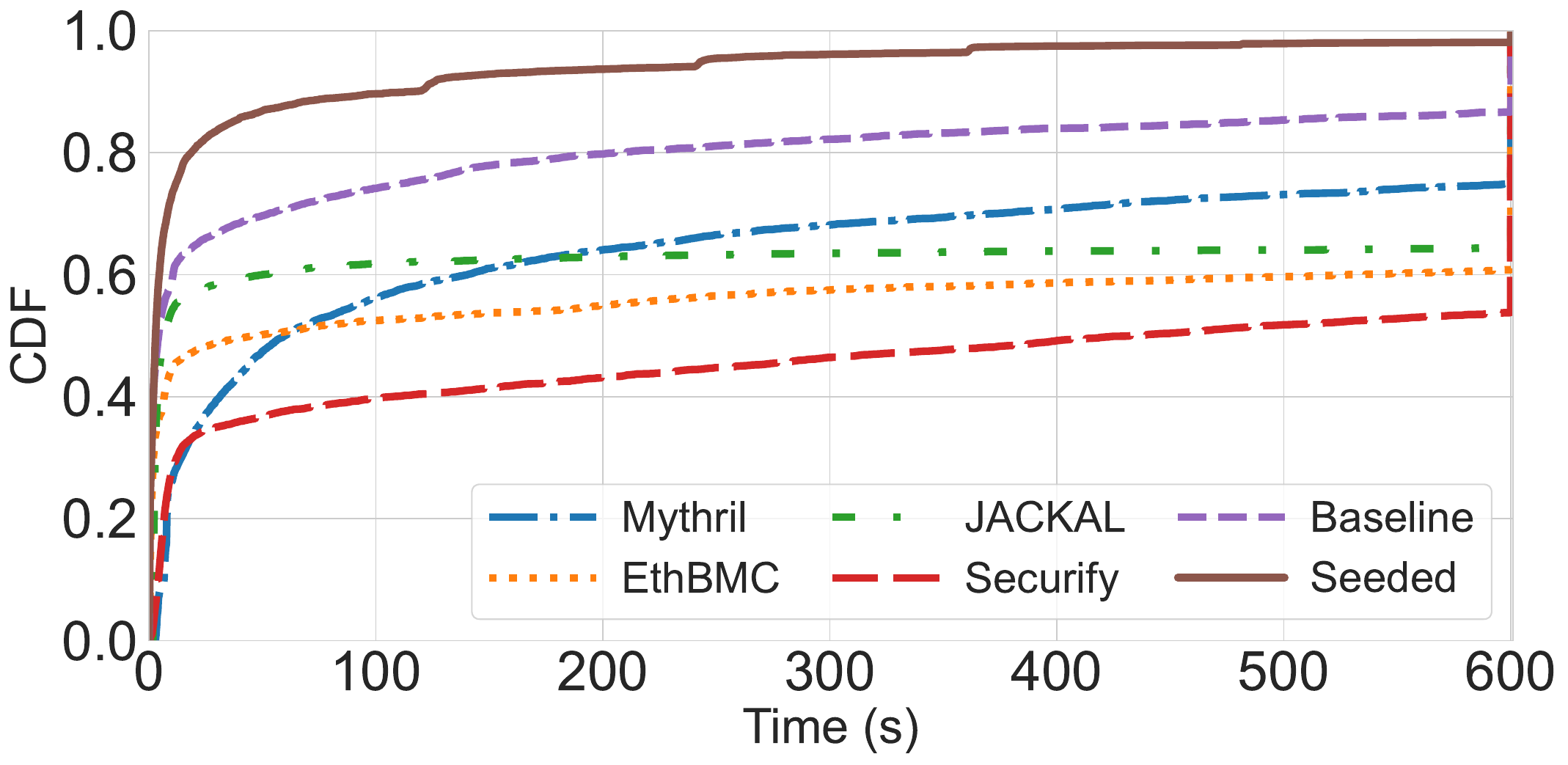}
    \caption{CDF of the time taken by \name, Mythril, ETHBMC, JACKAL, and Securify to analyze the smart contract.}
    \Description{A cumulative distribution plot comparing analysis time across six tools: Seeded, Baseline, Mythril, ETHBMC, JACKAL, and Securify. The horizontal axis shows time in seconds from 0 to 600, and the vertical axis shows the cumulative fraction of analyzed contracts. The Seeded curve rises fastest and remains highest throughout the range, while Securify and ETHBMC increase more slowly.}
    \label{fig:performance}
\end{figure}

Beyond its effectiveness in identifying vulnerabilities, we further evaluate \name's runtime.
As shown in~\autoref{fig:performance}, \name achieves shorter analysis time in the seeded mode than in the baseline mode---it can analyze about 90\% of the dataset's contracts within 100 seconds. This demonstrates the effectiveness of seeded symbolic execution. 
For other tools, our runtime advantage may result from multiple factors. For example, these tools were originally designed with different goals, and the additional analysis overhead from checking issues such as unprotected \texttt{SELFDESTRUCT} may affect their performance.

\done%

\subsubsection{Exploit generation and loss estimation}
For each contract with identified vulnerabilities, we generate and validate exploits, and compute the loss for realized exploits.
In total, we successfully exploit \confirmedbugs out of \identifiedbugs vulnerable contracts through local simulation.
This can be compared to state-of-the-art tools such as JACKAL, which only succeeded in exploiting 31 of the 529 contracts it flagged as vulnerable \cite{gritti2023confusum}.
Indeed, previous work highlights that even when an exploit cannot be automatically generated, the input code is not necessarily safe.
Rather, that may be because generating the correct inputs to trigger the vulnerability at hand may be highly time consuming.
Thus, prior work recommends manual analysis when encountering such cases in practice \cite{gritti2023confusum}.
This shows that automatic exploit generation, when successful, serves to lower the amount of manual work when compared to using tools which lack this capability altogether,
like Mythril \cite{mythril}, Securify~\cite{tsankov2018securify}, and Slither~\cite{feist2019slither}.

\parhead{Exploit synthesis limitations}
To better understand why exploits are sometimes not generated for contracts labeled as vulnerable, we manually inspect 100 randomly selected contracts.
In 65 cases, this is due to multiple \texttt{CALL} instructions along the exploit path that require precise parameter constraints.
The remaining 35 cases likely stem from Greed's inability to generate correct parameter constraints.
The complexity of generating parameter constraints is thus a target for future work.

\parhead{Component contribution to verified exploits}
We further break down which components of \name enable the 394 verified exploits. Among them, 178 exploits require control-flow deobfuscation to recover an analyzable CFG, and 209 involve \textit{risky} but not fully controllable \texttt{CALL}s that nonetheless execute successfully in our local validation environment. In addition, historical transactions help identify 308 potentially vulnerable contracts that are missed by the full-symbolic baseline; due to the constraints of our exploit synthesis strategy, 24 of these result in verified exploits.

Beyond component-level ablation, we analyze the properties of the \confirmedbugs exploits themselves. Among these exploits, 141 require the attacker to control \texttt{tx.origin}, for example through phishing-style attacks, while the remaining exploits do not rely on such control.

\parhead{Loss estimation}
We further calculate a lower bound on the potential loss if the synthesized exploits were carried out.
Since the balance of the victim contract changes over time, we use two reference points: the balances at the time of writing (June 5th, 2025), and the historical high balances (including vulnerable historical versions; see~\autoref{sec:historical-versions}), which represent a worst-case attack.
For simplicity of comparison, we use the token price on June 5th, 2025, to convert the loss to USD.
We focus on seven major ERC-20 tokens: WETH, WBTC, USDC, USDT, DAI, UNI, and LINK, so our results are a lower bound.
If a contract's vulnerability is limited to specific tokens, we only consider the affected tokens.

The results are shown in \autoref{tab:total-loss}. The ``current'' and ``maximal'' columns refer to the current balances and historical high balances, respectively.  
As shown in the table, the total estimated economic loss caused by the synthesized exploits is about 10.6 million USD.
Among them, WETH accounts for the largest portion, with an estimated loss of 7.6 million USD.
These losses suggest that profitable MEV bots are vulnerable and exposed to substantial financial risks.

\begin{table}[htbp]
    \centering
    \caption{A conservative estimate of the potential loss if the synthesized exploits were carried out.}
    \resizebox{\linewidth}{!}{
    \begin{tabular}{l|rr|rr}
        \toprule
\multirow{2}{*}{\textbf{Token}} & \multicolumn{2}{c|}{\textbf{Current}} & \multicolumn{2}{c}{\textbf{Maximal}} \\
\cmidrule{2-3} \cmidrule{4-5}
              & \textbf{Amount} & \textbf{Loss (\$)} & \textbf{Amount} & \textbf{Loss (\$)} \\
        \midrule
        WETH & 114.4 & 299,626.9 &  2,895.1 & 7,585,325.4 \\
        WBTC & 0.2 & 20209.0 & 20.0 & 2,100,271.7 \\
        USDC & 55,730.7 & 55,730.7 & 296,716.3 & 296,716.3\\
        USDT & 9,441.1 & 9,441.1 & 282,295.5 & 282,295.5 \\
        DAI& 50,386.6 & 50,386.6 & 185,609.3 & 185,609.3 \\
        UNI & 16,131.7 & 101,952.6 & 16,131.7 & 101,952.6 \\
        LINK & 2,076.4 & 28,799.4 & 3,061.2 & 42,458.8 \\
        \midrule
        \textbf{Total} & --- & \textbf{566,146.3} & --- & \textbf{10,594,629.6} \\
        \bottomrule
    \end{tabular}
    }
    \label{tab:total-loss}
\end{table}

\begin{tcolorbox}[title=Answer to \textbf{RQ-2}, left=2pt, right=2pt, top=2pt,bottom=2pt]
\name identifies \identifiedbugs potentially vulnerable MEV bot contracts and successfully generates effective exploits for \confirmedbugs of them, with a potential loss of \$10.6M. Its vulnerability detection outperforms other tools.
\end{tcolorbox}

In closing, we would like to clarify that the limitations or bugs in Greed and Gigahorse should be considered separately from the evaluation of \name. \name itself can be implemented on top of any existing tool with some additional engineering effort. The goal of this evaluation is to show that existing tools fail to address control flow obfuscation and suffer from performance issues on complex contracts. More importantly, they miss a significant number of asset management vulnerabilities. When integrated with \name, these tools can identify more potential vulnerabilities that would otherwise be missed.

\subsection{Attacks in the Wild}
\label{sec:evaluation-attacks}
Our previous evaluation confirms that \name can identify {\em new} vulnerabilities not reported or exploited in practice. In this section, we collected data on {\em real-world attacks} to (1) evaluate whether \name can discover vulnerabilities in attacked smart contracts; (2) quantify the potential loss that could have been saved by \name; (3) understand the prevalence of real-world exploits that target asset management vulnerabilities in closed-source smart contracts.

\parhead{Detecting attacks}

We start with 65,758,934 Ethereum transactions sent to \numberofcontracts MEV bot contracts from January 2021 to May 2025, from which we aim to identify attack transactions.
We note that our goal is not to conduct a comprehensive measurement study (in particular, we do not aim to be exhaustive). Thus, we employ a relatively simple method supplemented by manual verification.

First, based on how MEV bots work, we observe that external calls from a strange smart contract that an MEV bot had no prior interaction with are likely attacks.
Thus, we narrow down to transactions involving strange callers, followed by an ERC-20 asset transfer from the bot's account (or approval to transfer), a necessary action to steal assets.
These two rules result in 164,404 attack candidates, which are still too numerous to be manually verified. We further narrow down by looking for a specific malicious pattern: callbacks from an unintended caller. E.g., \texttt{uniswapV3SwapCallback} is intended to be called by Uniswap, and calls by others are likely an attack attempt.
Finally, this brings the number down to 104.

We verify all of them by checking whether the assets were transferred to the attacker's accounts, following a strict rule to determine if an account belongs to the adversary: we consider an account to belong to the adversary if it has no prior interaction with the victim contracts, and almost all of its transactions are related to the attack, and created malicious contracts explicitly for the attack. All results have been cross-validated by multiple authors to ensure accuracy.

We confirmed that all 104 detected transactions were adversarial, involving 50 malicious contracts and 36 victim MEV bot contracts.
All attacks involve the attacker bypassing \texttt{tx.origin} checks, supporting the assumption made in~\autoref{sec:vulnerability-detection} that origin-based access control can be bypassed under realistic attack conditions in MEV supply chains.
Moreover, all vulnerable contracts exploited by attackers are closed-source, and 14 of them are obfuscated; identifying vulnerabilities in these obfuscated contracts requires our deobfuscation step and would likely be missed prior to exploitation.

Our method is tailored to specified patterns and thus has no false positives, but may miss certain attacks.
Without ground truth, one cannot evaluate false negatives.
To the best of our knowledge, this is one of the largest real-world datasets of validated phishing-based exploits targeting asset management vulnerabilities in MEV bots.

\parhead{\name's ability to re-discover attacks}
We apply \name to victim contracts and find that it successfully identifies 27 as vulnerable.
If the searchers had used \name, most of the attacks could have been prevented.
In contrast, Mythril, ETHBMC, and JACKAL do not flag these victim contracts as vulnerable.

For the nine vulnerable smart contracts not detected by \name, four are due to failures in Gigahorse---similar to the issue described in~\autoref{sec:vulnerability-detection-evaluation}---and the remaining five are caused by Greed, which incorrectly classifies vulnerable paths as infeasible. These findings suggest that, if Gigahorse and Greed functioned as intended, \name could potentially detect all of these vulnerabilities.

\parhead{Victim loss} The earliest observed attack occurred in July 2021, resulting in a loss of 30 ETH (\$76K), while the most recent attack took place in April 2025. The single largest attack caused a loss of 250 ETH (\$636K).
The total losses from these incidents amount to about \$2.76M. The losses are calculated using the token price at the time of each attack.
We notice that only three of these attacks have been previously reported.
Our analysis indicates that \$2.45M of the total loss could have been mitigated if \name had been used by searchers before these incidents. This highlights that vulnerabilities in asset management smart contracts remain an ongoing security concern and often go unnoticed by developers.

\begin{tcolorbox}[title=Answer to \textbf{RQ-3}, left=2pt, right=2pt, top=2pt,bottom=2pt]
We identify 104 MEV phishing attacks in the wild, causing about \$2.76M in losses to searchers.
Only three of them have been publicly reported.
\name detects asset management vulnerabilities in most of the exploited contracts and could have prevented losses totaling \$2.45M.
\end{tcolorbox}

\section{New Attack Pattern: MEV Phishing Attack}
In this section, we delve into the attacks identified in~\autoref{sec:evaluation-attacks} to understand how they occurred and to show that our vulnerability oracle matches real-world exploitation patterns.

As in the ``Destroyer Inu'' attack (\autoref{sec:destroyer-inu}),
the adversary in each incident creates a special MEV opportunity
to lure the victim into interacting with an attacker-controlled contract. Once searchers try to capture this MEV opportunity via their MEV bot, they fall into the phishing trap, as the malicious contract is now embedded in their MEV supply chain, enabling a malicious \texttt{CALL} with adversary-controlled or risky parameters.
We refer to these as \textit{MEV phishing attacks} and further categorize them into two types based on the attack vectors: \textit{token-based} and \textit{pool-based} MEV phishing attacks, comprising 101 and 3 incidents, respectively.
We provide a detailed analysis of both attack types in~\autoref{sec:additional-mev-phishing-attacks}. 

Beyond malicious tokens and pools, we identify a distinct class of attacks originating from other components in the MEV supply chain. To investigate such attacks, we modify existing heuristics to detect components that, in principle, should not invoke the bot contract but do so in practice.
This led us to identify the refund address---a component that, in principle, should only receive refunds from searchers in exchange for providing MEV opportunities~\cite{bloXrouteBackRunMe}, and that should not interact with the bot contract. We refer to this class as \textit{refund-based} attacks.
To provide a better understanding of how real-world attackers exploit asset management vulnerabilities through sophisticated MEV phishing strategies, we present a case study of refund-based attacks. This novel attack vector is exemplified by a real-world transaction \href{https://etherscan.io/tx/0x26361798094d7532c0b8dfbed4c857265c66391040eef07f91fafcd420d47df0}{0x263$\dots$df0}.

\begin{figure}[thbp]
    \centering
    \includegraphics[width=\linewidth]{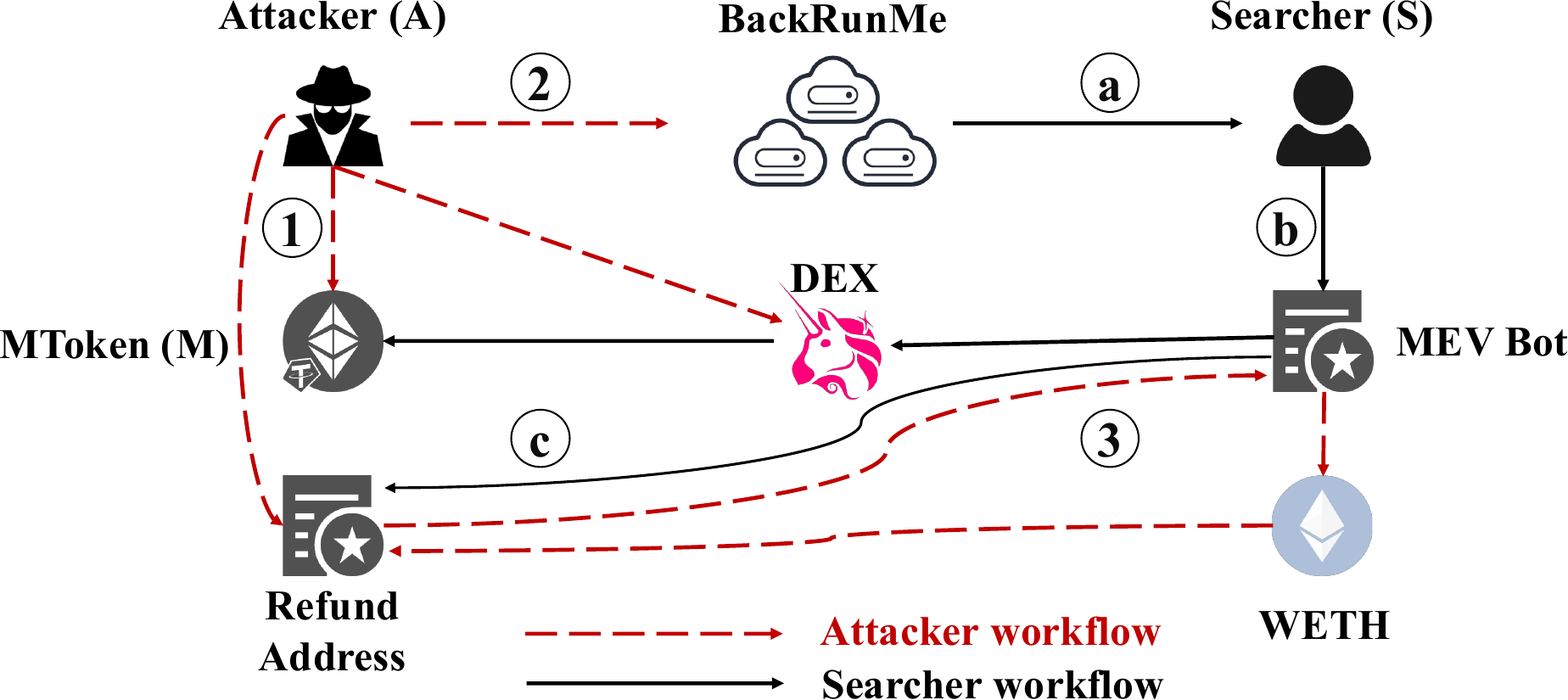}
    \caption{Refund-based MEV phishing attack workflow.
    \Description{A workflow diagram showing two interacting execution paths in a refund-based MEV phishing attack. Red dashed arrows represent the attacker workflow and black solid arrows represent the searcher workflow. The attacker first creates a malicious token, then submits bait transactions through BackRunMe. The searcher forwards the opportunity to an MEV bot, which executes trades through a decentralized exchange. During execution, value is redirected through a malicious refund address, and WETH is transferred along the attacker path.}
    }
    \label{fig:mev-phishing-attack}
\end{figure}

\parhead{Refund-based MEV phishing attack}
Most MEV phishing attacks rely on malicious tokens as attack vectors.
These tokens store specially crafted calldata associated with specific target accounts.
When a transaction originates from that targeted account, the token automatically triggers the attack against the MEV bot contract using the prepared calldata.
However, this method has a key limitation: after repeated targeting, searchers may detect the issue and adopt countermeasures, such as validating that tokens strictly follow the ERC-20 standard.
A recent variant departs from using malicious tokens and instead exploits the mechanics of MEV refund services. The attack flow is illustrated in~\autoref{fig:mev-phishing-attack}.

The attacker first deploys a malicious contract as the refund address (\textcircled{1}) and then creates an MEV opportunity, involving a swap, which is submitted to the BackRunMe service~\cite{bloXrouteBackRunMe} (\textcircled{2}). The tokens in the swap follow the ERC-20 standard, avoiding detection mechanisms targeting non-compliant tokens.

A searcher receives the opportunity information via BackRunMe (\textcircled{a}) and invokes its MEV bot to extract it (\textcircled{b}). As part of the BackRunMe protocol, the searcher must send a refund to the attacker-specified address (\textcircled{c}). Upon receiving the refund, the malicious contract is triggered and calls the MEV bot contract using carefully crafted calldata (\textcircled{3}).
Since the transaction originates from the searcher, it bypasses origin-based access control and enables unauthorized token transfers to the attacker.

\section{Related Work}
\label{sec:related-works}

\parhead{Static analysis}
Static analysis is a method of examining code without executing it to detect software vulnerabilities.
Static analysis tools for smart contracts, such as Vandal~\cite{brent2018vandal}, Securify~\cite{tsankov2018securify}, MadMax~\cite{grech2018madmax}, Slither~\cite{feist2019slither}, and Ethainter~\cite{brent2020ethainter}, typically aim for completeness but often produce false positives, requiring manual verification.
Moreover, a well-known challenge for static analysis is control flow obfuscation~\cite{moser2007limits}, which often prevents them from constructing accurate control flow graphs, thus reducing effectiveness.
Specifically, vulnerable logic may be treated as unreachable, leading to false negatives.
The deobfuscation technique in \name complements these tools by improving accuracy and efficiency.

\parhead{Symbolic execution}
Symbolic execution explores program execution paths by treating inputs as symbolic variables rather than concrete values.
Tools like Oyente~\cite{luu2016making}, Mythril~\cite{mythril}, teEther~\cite{krupp2018teether}, Manticore~\cite{mossberg2019manticore}, and ETHBMC~\cite{frank2020ethbmc} apply this approach to smart contracts to identify issues like reentrancy and integer overflows.
However, like static analysis, symbolic execution struggles with symbolic jump destinations caused by control flow obfuscation.
The de-obfuscation module of \name allows it to handle contracts with obfuscated control flow and detect vulnerabilities.

\parhead{Fuzzing}
Fuzzing is a dynamic testing technique that mutates inputs to uncover software vulnerabilities.
For smart contracts, fuzzing applies random and unexpected inputs, such as crafted calldata and reentrant calls, to test contract behavior.
Tools developed for EVM smart contracts include ContractFuzzer~\cite{jiang2018contractfuzzer}, Echidna~\cite{grieco2020echidna}, Smartian~\cite{choi2021smartian}, Confuzzius~\cite{torres2021confuzzius}, ItyFuzz~\cite{shou2023ityfuzz}, and MAU~\cite{chen2024towards}.
However, while state-of-the-art fuzzing tools have improved the mutation efficiency, they remain insufficient for our problem.
The vulnerability logic we focus on is often triggered only by specifically crafted calldata, which differs across contracts.
As a result, fuzzers may treat these inputs as arbitrary byte mutations, making the exploration process highly inefficient.

\parhead{Attacks against searchers}
When searchers send transactions to the public mempool, their transactions and execution logic become visible to adversaries, enabling real-time imitation and front-running.
Qin et al.~\cite{qin2023blockchain} generalize this attack through Ape, a framework that uses dynamic program analysis to automatically generate adversarial contracts.
Interestingly, the same technique can defend against malicious searchers. Zhang et al. propose STING, a defense mechanism that identifies attacking transactions and instantly synthesizes counterattack smart contracts~\cite{zhang2023your}.

However, as private mempools become more common~\cite{yaish2023suboptimality,yang2024decentralization}, previous solutions lose effectiveness because searchers' pending transactions are invisible when sent to a private mempool.
Various works attempt to circumvent these limitations.
The GhostTX attack by Yaish et al.~\cite{yaish2024speculative} tricks searchers into bundling adversarial transactions that appear profitable but are invalid, lowering their standing in reputation-based private mempools.
Shou et al.~\cite{shou2024backrunner} introduce BACKRUNNER, which exploits (1) the delay between exploit deployment and execution, and (2) incomplete asset drainage, enabling backrunning to recover funds.
These works target searchers by analyzing their logic and competing for the same MEV.
In contrast, we exploit vulnerabilities within searchers' contracts directly.

\parhead{MEV attacks and defenses}
Although MEV searchers are considered victims in this paper, prior work often considers them attackers.
They perform various MEV activities for profit~\cite{qin2022quantifying,li2023demystifying,ferreira2024rolling}, including front-running~\cite{torres2021frontrunner}, sandwiching~\cite{zhou2021high}, liquidations~\cite{qin2021empirical}, and arbitrages~\cite{zhou2021just,mclaughlin2023large,heimbach2024nonatomic,oz2025pandora}. Some of these activities, notably front-running and sandwich attacks, constitute direct attacks on users by manipulating their transaction order to extract value. To mitigate the negative effects of MEV on users, both academia and industry have proposed various countermeasures, including private transaction channels~\cite{yang2024sok,yang2024decentralization}, time-based order fairness protocols~\cite{kelkar2020order,zhang2020byzantine,kelkar2023themis}, and front-running-resistant AMM designs~\cite{zhou2021a2mm,wadhwa2024data,zhang2024rediswap}.

\parhead{Blockchain phishing}
Prior research on blockchain phishing has mainly focused on attacks against ordinary wallet users.
These studies examine several common attack vectors, including phishing websites~\cite{he2023txphishscope}, deceptive transaction payloads~\cite{chen2025dissecting}, address poisoning~\cite{ye2024interface,guan2024characterizing,tsuchiya2025blockchain}, and phishing contracts~\cite{he2025phishing}, all of which aim to induce users to sign malicious transfers, approvals, or other harmful on-chain interactions.
In contrast, the attacks we study target MEV bots: attackers construct seemingly profitable opportunities that exploit bot-specific trading logic and trigger automated execution.
By embedding a malicious contract into the induced execution path, they cause the MEV bot to invoke the contract during execution, thereby satisfying access-control conditions that would not hold under direct attacker interaction.

\section{Discussion}

\subsection{Current Scope and Extension}

Our evaluation focuses on asset management vulnerabilities, a particularly important class in smart contracts because they can lead to substantial potential and realized losses, as shown in our vulnerability detection and validated attacks (\autoref{sec:vulnerability-detection-evaluation} and \autoref{sec:evaluation-attacks}). Within this scope, \name currently expresses vulnerability patterns as reachable \texttt{CALL} instructions whose critical parameters are either adversary-controllable or fixed but risky, for example, a controllable call target or transfer recipient, or a fixed ERC-20 token contract or token-transfer function. Within this scope, our current implementation focuses on \texttt{CALL}, which underlies the dominant asset-movement mechanism in the contracts we study, including ERC-20 token transfers, and captures the majority of real-world exploits in our dataset.

Other call-related opcodes, namely \texttt{CALLCODE}, \texttt{STATICCALL}, and \texttt{DELEGATECALL}, have different execution behavior and expose different vulnerability patterns, so the vulnerability oracle and taint propagation rules must be adapted accordingly. For example, \texttt{STATICCALL} cannot directly modify state, while \texttt{DELEGATECALL} and \texttt{CALLCODE} execute external code in the caller's context and are therefore more naturally associated with patterns such as storage collision~\cite{ruaro2024not}. Extending \name to these call-related opcodes would therefore expand its vulnerability coverage beyond the current scope. This can be done by adding opcode-specific execution modeling, rules for
taint propagation, and corresponding updates to the vulnerability oracle. We leave these extensions to future work.

\subsection{Tradeoff Between Cost and Security}
\label{sec:tradeoff-between-cost-security}
Unlike traditional programs, computation on EVMs is expensive: users must pay high fees for complex transactions.
This implies that, if rigorously enforced, access control can be costly.
As illustrated in~\autoref{fig:check-pool-address} (in \autoref{sec:uniswap-v3-access-control}), access control for a Uniswap V3 pool requires computing the expected address of the pool according to the \texttt{CREATE2} specification~\cite{buterin2018eip1014}, which consumes 449 gas per interaction.
Assume that each transaction sent to the MEV bot involves an interaction with a Uniswap V3 pool, requiring one verification.
If a searcher sends 10K transactions to the MEV bot contract per month, the total gas usage amounts to 4.5M.
In contrast, if the contract employs an \textit{incorrect but cheap} verification by checking \texttt{tx.origin}, the total gas usage is reduced to 80K.

The cost could even be zero if the contract does not employ any access control mechanism. For example, \name also identified a \href{https://etherscan.io/address/0x64F2095CC11e4726078F4A64d4279c7e7fB7e6Ec}{1inch contract} without access control to protect its asset management logic, allowing any account to transfer ERC-20 tokens from its account.
We disclosed this issue to the 1inch team.
Their response indicated that the contract only holds residual tokens from executed transactions and does not affect user-owned assets.
In this case, the cost of a rigorous access control mechanism seems less justified, considering that the exploit only extracts residual tokens. It may be more reasonable to save gas for legitimate users of the contract.

This raises an interesting question: What is the best tradeoff between cost and security? Many profit-driven smart contracts may be sensitive to cost, while a rigorous approach would increase transaction costs. Therefore, a promising direction is designing systems with strong security and low cost.

\subsection{Benefits to Developers}
The evaluation in~\autoref{sec:evaluation} shows that \name can detect vulnerable smart contracts and thus help prevent attacks. Beyond this empirical result, \name is particularly useful in development and auditing settings where bytecode-level analysis is necessary. This includes contracts written directly in low-level bytecode, for example, to implement control-flow patterns such as indirect jumps that Solidity does not natively support, as is common in some MEV bots~\cite{degatchi2023smart,degatchi2024mev}. It also includes contracts developed with alternative languages or custom toolchains that lack mature security analysis support, as well as closed-source contracts for which source code is unavailable. In such settings, analyzing deployed bytecode directly provides assurance about the actual artifact executed on chain, including cases where obfuscation or low-level transformations may introduce unintended vulnerabilities.

In addition, developers and auditors are often in a stronger position than attackers to benefit from \name's seeded symbolic execution. Before deployment, or immediately after deployment, they can supply representative transactions derived from intended usage and test critical execution paths systematically. By contrast, an attacker is limited to the transactions that are already observable on chain, which restricts the effectiveness of the same analysis strategy.

\done%

\section{Conclusion}

We have presented \name, an EVM bytecode analysis tool optimized for closed-source smart contracts such as MEV bots.
\name can effectively deobfuscate control flow and identify asset management vulnerabilities, an ability that existing tools do not offer. 
\name also employs transaction-seeded symbolic execution to improve efficacy.
We evaluated \name against real-world MEV bots, using historical transactions to facilitate symbolic execution.
Among \numberofcontracts MEV bots we studied, \name detects vulnerabilities in \identifiedbugs of them.
Further, \name automatically generates exploits against \confirmedbugs of them, with a potential loss exceeding \totalloss.
We also discovered 104 attacks in the wild that exploited asset management vulnerabilities against 36 MEV bots, resulting in a total loss of \$2.76M.

\begin{acks}
The authors thank the anonymous reviewers and the shepherd for their constructive comments and guidance.
The authors also thank the greed project team in the Computer Security Group at UCSB for answering our questions.
The authors used ChatGPT to check grammar and typos.
\end{acks}

\bibliographystyle{ACM-Reference-Format}
\bibliography{ref}

\appendix

\section{Empirical Study on Top Ethereum Smart Contracts}
\label{sec:ethereum-smart-contracts}

To identify the most active smart contracts on Ethereum, we rank them by the number of transactions that have interacted with them since 2021.
Our dataset includes the top 50K smart contracts, with at least 899 transactions sent to each contract.
For each smart contract, we check whether it is open-source on Ethereum by querying the Etherscan API~\cite{etherscan2025docs}.
Among these 50K smart contracts, 7,328 (15\%) are closed-source. Among the closed-source contracts, 4.2\% are obfuscated (as indicated by code coverage below 90\%).

To better understand the importance of closed-source contracts, particularly those with obfuscation, we query each contract's ETH balance and ERC-20 token balances at block height 23,149,884 (August 16, 2025). We convert these holdings into USD using the corresponding daily token prices. We then apply the static analysis tool Gigahorse~\cite{grech2019gigahorse,grech2022elipmoc,lagouvardos2024incredible} to each closed-source smart contract's bytecode and measure code coverage as the percentage of basic blocks that are reachable according to the analysis.
In the end, we find that contracts with obfuscation (identified as those with code coverage below 90\%) hold over \$37M worth of crypto assets (ERC-20 tokens and ETH), which highlights the importance of understanding their security risks. 

\section{Case Study: the Destroyer Inu Attack}
\label{sec:attack-analysis}
\label{sec:destroyer-inu}

A notable attack on a closed-source smart contract that relies on external inputs for its operation took place on July 1st, '24, causing a loss of 22 ETH (worth \$51,056 at the time)~\cite{0xprincess2024tweet}.
As the attack involves an attacker-created ERC-20 token called ``Destroyer Inu'', we use the same name for the attack itself.
This case study shows that control-flow obfuscation adopted by the victim contract ends up hiding vulnerabilities from analysis tools---without stopping real-world attackers---arguably making contracts less secure.

\parhead{tx.origin phishing attack}
The first issue is a classic vulnerability where \texttt{tx.origin} is compared to a hard-coded trusted address.
This vulnerability is known as the \textit{tx.origin phishing attack}~\cite{soliditybyexample2025origin}, dating back to at least 2016~\cite{RemovetxoriginIssue683}.
In such attacks, an attacker tricks the victim into calling a malicious contract, which then calls the victim's contract; since both calls have the same \texttt{tx.origin}, the malicious smart contract can execute the victim smart contract with the same privileges as the victim herself. 

A more severe issue that follows the incorrect access-control check is improper asset management. However, existing tools miss the full exploit because of control-flow obfuscation. Existing tools such as Mythril flag it only as a ``low-risk'' issue. We will discuss the full vulnerability next.

\begin{figure}[tbp]
\centering
\begin{lstlisting}[style=YulStyle,xleftmargin=5.0ex]
assembly {
entry 0x0:
    let from := origin() // tx.origin
    let expected := 0xdead...beef
    // access check
    if iszero(eq(from, expected)) {revert(0, 0)}
    // read two bytes from calldata
    // as the jump destination.
    // suppose it's 0x0a00...
    let data := calldataload(0x84)
    let dst := shr(0xf0, data)
    jump dst      // jump 0x0a00
    
jumpdest 0x0a00:
    let token := calldataload(0x86)
    let to := calldataload(0xa6)
    let value := calldataload(0xc6)
    let ptr := mload(0x40)  
    mstore(ptr, 0xa9059cbb) // ERC-20 transfer
    mstore(add(ptr, 4), to)  
    mstore(add(ptr, 36), value)
    // call target.transfer(to, value)
    let success := call(
        gas(), token, 0, ptr, 68, 0, 0
    )
    if iszero(success) { revert(0,0) }  
    
jumpdest 0x0b00: ...
}
\end{lstlisting}
\caption{Pseudo-Yul MEV bot smart contract. The code starting at \texttt{0x0a00} contains a vulnerable function call whose input is determined by the calldata, allowing an adversary to transfer any amount of ERC-20 tokens from the MEV bot's account to their own. Note that {\color{blue} \texttt{jump} } and {\color{blue} \texttt{jumpdest}} are not part of Yul syntax but are included for better readability.
\Description{A pseudo-Yul code listing of an MEV bot smart contract. The contract first checks tx.origin against a fixed address, then derives a jump destination from calldata and jumps to it. At jump destination 0x0a00, calldata determines the token address, recipient, and transfer amount used in an ERC-20 transfer call, illustrating how externally supplied calldata controls the vulnerable asset-transfer operation.}
}
\label{fig:destroyer-inu}
\end{figure}

\parhead{tx.origin phishing attack, obfuscated}
In the Destroyer Inu Attack, the victim contract is obfuscated. We use pseudo-Yul code to illustrate the smart contract's logic in \autoref{fig:destroyer-inu}. 
First, at the entry point of the MEV bot's smart contract in lines 3 to 6, it implements an access control mechanism based on the transaction's origin to prevent unauthorized addresses from interacting with the contract.
Then, in lines 10 to 12, the contract proceeds by jumping to a specific place in code, as determined by the calldata supplied by the transaction.
This is unlike ``standard'' smart contracts that match function calls using the so-called \emph{function selector} defined as the first 4 bytes of the calldata~\cite{solidity2025abispec}.
In this MEV bot smart contract specifically, when the two calldata bytes at positions \texttt{0x85} and \texttt{0x86} are \texttt{0x0a00}, the program jumps to the code segment starting at \texttt{jumpdest 0x0a00}.  
The functionality of this segment is to manage the transfer of a specific amount of an ERC-20 token to a target address.
Therefore, it loads the address of the ERC-20 token, the recipient address, and the amount from the calldata (lines 15 to 17).
Then, in lines 18 to 25, it constructs the input based on these parameters and invokes the ERC-20 transfer.

\parhead{Exploit}
To an attacker who can remove control-flow obfuscation, the vulnerability is immediate: the victim contract allows transferring any amount of any ERC-20 tokens held by this contract to another address if an attacker can bypass the tx.origin check via many forms of \emph{tx.origin phishing}, which is precisely what happened. 

\begin{figure}[htbp]
    \centering
    \includegraphics[width=\linewidth]{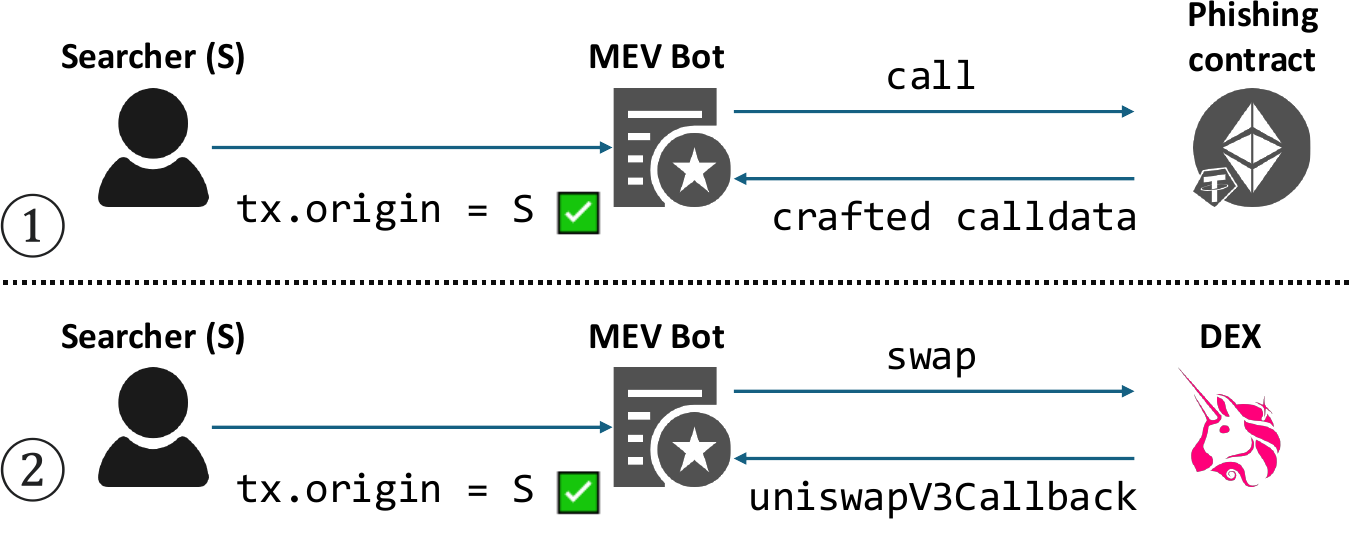}
    \caption{Two scenarios where an MEV bot contract receives a callback: a phishing contract (\ding{172}) and a DEX (\ding{173}).}
    \Description{A comparison of two callback paths involving an MEV bot. The upper scenario shows the bot calling a phishing contract and receiving crafted calldata. The lower scenario shows the bot interacting with a decentralized exchange during a swap and receiving a uniswapV3Callback. Both scenarios begin with a searcher transaction, and both indicate that tx.origin equals the searcher address.}
    \label{fig:analysis-of-two-cases}
\end{figure}

\parhead{Attack analysis}
A direct cause of the Destroyer Inu attack is that the victim's contract relies on a vulnerable comparison to block unintended calls by adversaries.
A strawman mitigation is to use a hard-coded comparison against \texttt{msg.sender} instead of \texttt{tx.origin}.
However, this solution does not meet real-world requirements when verifying if the caller is the searcher (the contract's creator and owner), as, in practice, the contract may be called by other different contracts.
For example, Uniswap V3 pools require any contract that calls their \texttt{swap} function to implement a function called \texttt{uniswapV3SwapCallback} \cite{IUniswapV3SwapCallback}, which is invoked by pools at the end of executing \texttt{swap} to ensure that the interacting contract pays the pool the tokens owed for the swap (as illustrated in~\autoref{fig:analysis-of-two-cases}, scenario~\ding{173})

Other DeFi protocols, such as Uniswap V2, Sushiswap, and AAVE, also adopt similar callback designs.
In these cases, a hard-coded comparison against \texttt{msg.sender} is not effective because the caller may be any contract deployed by these protocols.
Compared to this, verifying \texttt{tx.origin} is not restricted by a specific caller and can prevent direct calls to the contract, but it introduces security risks.
Therefore, more precise protection is needed to prevent potential adversaries, which can be a high requirement for contracts, as real-world attack incidents have shown.

\section{Other Obfuscation Techniques}
\label{sec:obfuscation-tools}

We identify four contract obfuscation techniques based on previous works~\cite{collberg1997taxonomy,yu2022bytecode,zhang2023bian,zhang2025following}, summarized as: 

\begin{itemize}
    \item \textbf{Layout obfuscation} typically aims to reduce information available to a human reader, e.g., by removing comments and renaming functions and variables~\cite{collberg1997taxonomy,zhang2023bian}.
    \item \textbf{Data flow obfuscation} obscures how data is processed and accessed within the smart contract~\cite{collberg1997taxonomy,zhang2023bian}. 
    \item \textbf{Control flow obfuscation} alters the execution path of the smart contract to make its logical flow difficult to follow~\cite{collberg1997taxonomy,ma2025surviving}, which is our focus.
    \item \textbf{Preventive transformations} aim to hinder reverse engineering by exploiting weaknesses in current decompilers and deobfuscators~\cite{collberg1997taxonomy}. 
    E.g., contracts can self-destruct after execution to render their bytecode unavailable, preventing future inspection~\cite{zhang2025following}.
\end{itemize}

We empirically verified that the layout and data flow obfuscation do not interfere with existing analysis tools.

Two mature obfuscation tools are available for EVM smart contracts~\cite{zhang2025following}: BOSC~\cite{yu2022bytecode} and BiAn~\cite{zhang2023bian}. Since BOSC cannot guarantee that the obfuscated contracts remain deployable, it is not suitable for our setting. Following prior work~\cite{zhang2025following}, which also only used BiAn due to BOSC's limitations, we adopt BiAn for our evaluation.

We collected vulnerable smart contracts from the Smart Contract Weakness Classification Registry~\cite{swcregistry} and applied BiAn to each contract to generate an obfuscated version. Specifically, we apply two types of obfuscation: layout obfuscation and data flow obfuscation. We did not apply control-flow obfuscation, since the current implementation in BiAn is non-functional~\cite{zhang2025following,xf972025BiAn}.
We then compile the original vulnerable contracts and the obfuscated ones, and apply Mythril to detect their vulnerabilities.

Interestingly, we found that for 65 of the vulnerable smart contracts, layout and data flow obfuscation do not prevent the vulnerabilities from being detected; instead, the obfuscation even introduces new ``Assert Violation'' issues~\cite{SWC-110}.
For the remaining two contracts, we observed that Mythril could still detect issues in the obfuscated contracts at the source code level, but not at the bytecode level. We speculate this might be related to compiler optimizations.
In summary, our findings indicate that existing analysis tools like Mythril can still work effectively on smart contracts with both layout and data flow obfuscation.

\section{Additional Details on Taint Tracking}
\label{sec:taint-tracking}

\parhead{Design philosophy}
At the bytecode level, Solidity's complex data structures (e.g., mappings, arrays, structs) are compiled into memory offsets, calldata offsets, or storage slots. Accordingly, our taint tracking does not require field- or pointer-level semantics at the source level. Instead, we track taint at the granularity of offsets and slots: if a tainted value contributes to a memory offset or storage slot, the corresponding access is marked tainted. 
This offset- and slot-based tracking naturally captures fine-grained dependencies in high-level data structures.

Our taint analysis focuses on explicit data flows. We propagate taint only through data dependencies in the EVM (stack, memory, and storage) and do not model control dependencies (e.g., tainted branch conditions). This design ensures that the analysis captures whether adversarial calldata bytes directly influence the target address, function selector, or other \texttt{CALL} arguments.

\subsection{Byte-level Propagation Model}
\name implements a \textit{byte-level} propagation model to maintain high analysis precision. We define the taint state of a 256-bit EVM word $v$ as a vector of taint tags:
\[ \tau(v) = \langle b_0, b_1, \dots, b_{31} \rangle, \quad b_i \in \mathcal{P}(\mathbb{N}) \cup \{ \perp \} \]
where $\mathcal{P}(\mathbb{N})$ denotes the power set of natural numbers. Each element $b_i$ represents the set of \texttt{calldata} byte-offsets that influenced the $i$-th byte of $v$, or $\perp$ if it is untainted. This power-set representation enables \name to track multiple simultaneous taint sources for each byte. Such fine-grained granularity allows \name to distinguish between multiple variables packed within a single word (e.g., an \texttt{address} and a \texttt{uint96}). To fully capture the contract's execution environment, we further let $\mathcal{M}$ and $\mathcal{S}$ denote the taint mappings for the memory and persistent storage domains, respectively.

\parhead{Propagation semantics}
\name updates the taint state based on Gigahorse's three-address code (TAC) IR~\cite{grech2019gigahorse}. \autoref{tab:taint-rules} summarizes the formal semantics for key instructions. For instructions loading external input, such as $\texttt{CALLDATALOAD}(p)$, the model initializes the taint vector by mapping each byte to its specific calldata offset: $\forall i \in [0, 31], b_i = \{p+i\}$. 
To handle variables merging from different control-flow paths in the Gigahorse IR, \name resolves $\Phi$-nodes by selecting the taint vector of the \textit{most recent writer} on the actual execution trace (denoted as $\tau(u_{active})$). 

\begin{table}
\centering
\caption{Byte-level taint propagation rules in \name.}
\label{tab:taint-rules}
\resizebox{\linewidth}{!}{
\begin{tabular}{ll}
\toprule
\textbf{Instruction} & \textbf{Taint Update Rule} \\ \midrule
$v := \texttt{CALLDATALOAD}(p)$ & $\tau(v) = \langle p, p+1, \dots, p+31 \rangle$ \\
 $\texttt{CALLDATACOPY}(\mathrm{dst}, \mathrm{src}, \mathrm{len})$ & $\forall i \in [0, \mathrm{len}-1]: \mathcal{M}(\mathrm{dst} + i) = \mathrm{src} + i$ \\ \midrule
$v := \texttt{SHL}(n, u)$ & $\tau(v)[i] = (i + n/8 < 32) ? \tau(u)[i + n/8] : \perp$ \\
 $v := \texttt{BYTE}(i, u)$ & $\tau(v)[i] = \tau(u)[i], \text{ others are } \perp$ \\
$v := \texttt{AND}(\mathrm{mask}, u)$ & $\tau(v)[i] = (\texttt{byte}(\mathrm{mask}, i) \neq 0) ? \tau(u)[i] : \perp$ \\
$v := \texttt{ADD/SUB/OR}(u, w)$ & $\tau(v) = \tau(u) \cup \tau(w)$ \textit{(element-wise union)} \\ \midrule
$\texttt{MSTORE}(\mathrm{off}, u)$ & $\forall i \in [0, 31]: \mathcal{M}(\mathrm{off} + i) = \tau(u)[i]$ \\
$\texttt{MSTORE8}(\mathrm{off}, u)$ & $\mathcal{M}(\mathrm{off}) = \tau(u)[31]$ \\
$v := \texttt{MLOAD}(\mathrm{off})$ & $\forall i \in [0, 31]: \tau(v)[i] = \mathcal{M}(\mathrm{off} + i)$ \\ \midrule
$\texttt{SSTORE}(\mathrm{slot}, u)$ & $\mathcal{S}(\mathrm{slot}) = \tau(u)$ \\
$v := \texttt{SLOAD}(\mathrm{slot})$ & $\tau(v) = \mathcal{S}(\mathrm{slot})$ \\ \midrule
$v := \Phi(u_1, \dots, u_n)$ & $\tau(v) = \tau(u_{\mathrm{active}})$ \\
\bottomrule
\end{tabular}
}
\end{table}

\parhead{Context-sensitive inter-procedural flows}
To support complex contract logic, \name tracks flows through internal function calls:
\begin{itemize}
    \item \textbf{Function Entry:} Upon \texttt{TAC\_Callprivate}, the engine establishes a mapping $\alpha$ between caller registers and callee arguments, initializing $\tau(\mathrm{Var}_{\mathrm{callee}}) \leftarrow \tau(\mathrm{Var}_\mathrm{caller})$.
    \item \textbf{Function Exit:} Upon \texttt{TAC\_Returnprivate}, the taint vectors from the callee's return variables are propagated back to the caller's assigned result registers.
\end{itemize}

\section{Statistics on the Branch Table}
\label{sec:branch-table-study}

\begin{figure}
    \centering
    \includegraphics[width=\linewidth]{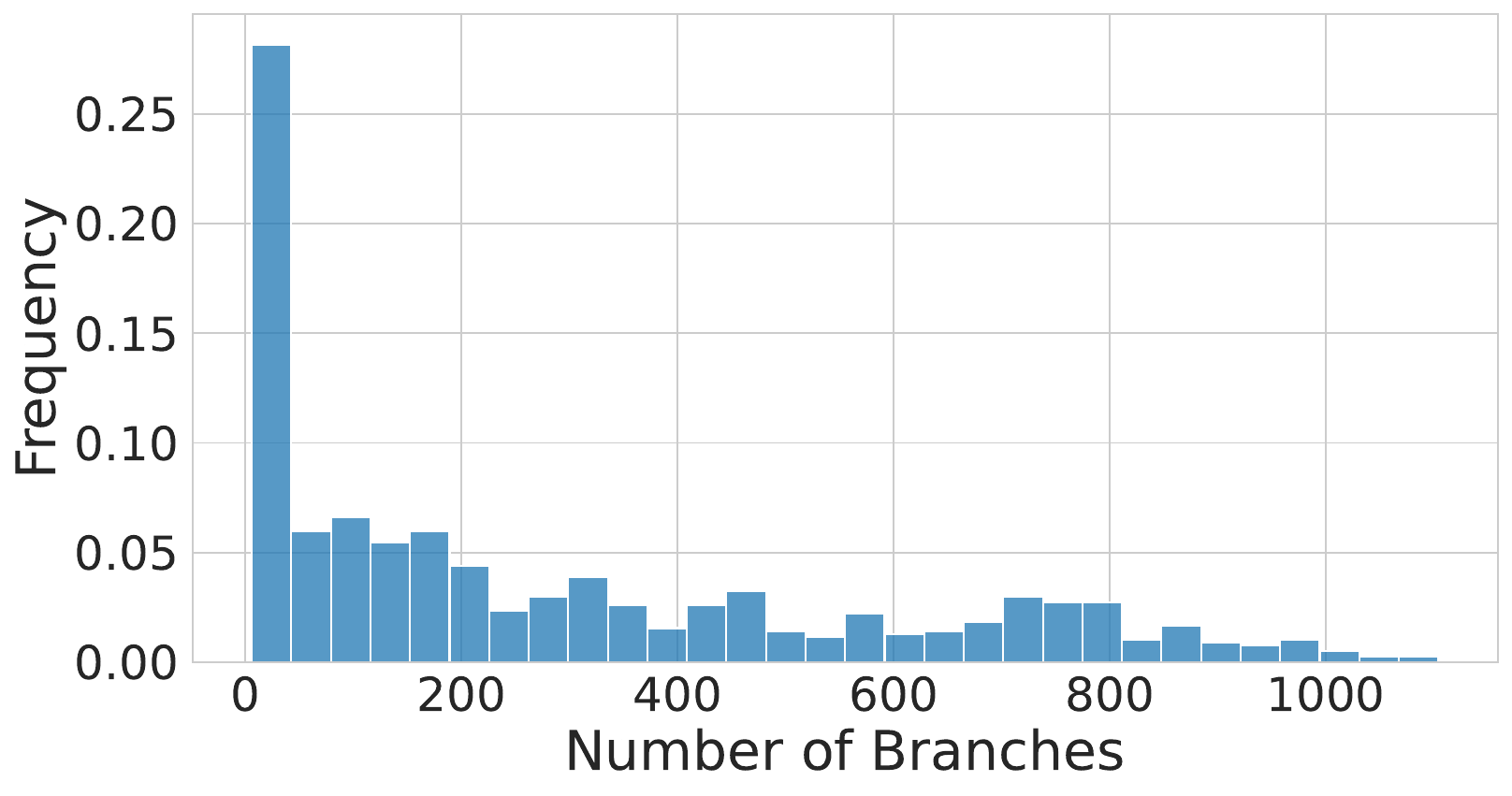}
    \caption{Distribution of branch counts in the \switchtable of each deobfuscated smart contract.}
    \Description{A histogram of branch-table sizes in deobfuscated smart contracts. The x-axis shows the number of branches per contract, and the y-axis shows the relative frequency. The highest concentration appears near small branch counts, with gradually decreasing frequency as the number of branches increases and a long tail reaching beyond 1000 branches.}
    \label{fig:branches-count}
\end{figure}

We count the number of branches in each inserted \switchtable for the obfuscated contracts, and the results are shown in~\autoref{fig:branches-count}.
It shows that 38\% of the contracts have fewer than 100 branches, and about 70\% have fewer than 400 branches, with a long tail extending to contracts containing over a thousand branches.
Overall, the majority of contracts contain relatively few branches in their branch tables, and we find that a high number of branches does not necessarily imply greater complexity, since none of the inserted \switchtable caused a timeout during our evaluation.

\section{Evaluation of the Pruning Rule}
\label{sec:pruning-rule}
To evaluate the effectiveness of the pruning heuristic, we conduct a comparative study by running \name in full symbolic execution mode both with and without pruning.
Specifically, the rule \textit{terminates any execution path} that visits the \switchtable more than twice.
We focus on contracts in our datasets (see~\autoref{sec:eval-setup}) that contain a \switchtable after deobfuscation.

Our evaluation shows that the pruning rule reduces state explosion in a small number of cases, enabling \name to complete symbolic exploration and identify vulnerabilities that would otherwise time out due to excessive path forking.
For the remaining contracts, the detection results are identical with and without pruning, in terms of both reachable vulnerable \texttt{CALL} instructions and verified exploits.
We did not observe any cases in which pruning suppresses vulnerability detection.
This is because repeated visits to the \switchtable usually reflect dispatcher-style control flow from obfuscation, while vulnerability-triggering paths do not require repeatedly traversing such logic within a single execution trace.

The pruning rule applies only to the fallback full symbolic execution mode.
When historical transactions are available, seeded symbolic execution follows concrete execution traces and does not use this heuristic.
As a result, the pruning rule does not affect the detection results reported in \autoref{sec:vulnerability-detection-evaluation}.
Overall, the pruning rule is a conservative optimization that improves scalability without degrading detection effectiveness on real-world obfuscated contracts.
Designing more precise mechanisms to control state explosion under heavy control-flow obfuscation remains future work.

\section{Historical Versions of MEV Bots}
\label{sec:historical-versions}

Although smart contracts on Ethereum are generally immutable once deployed, their bytecode at a given address may change over time through the use of \texttt{SELFDESTRUCT} and \texttt{CREATE2} opcodes~\cite{wood2014ethereum}. Specifically, \texttt{SELFDESTRUCT} removes the contract code from the state, and \texttt{CREATE2} allows redeployment of a new contract at the same address using a fixed deployer address, salt, and modified initialization code. 
As a result, a smart contract may have different bytecode at different points in time~\cite{ma2023abusing}. 

We define a \textit{version} of a smart contract as the specific bytecode deployed at a given address during a particular period. If a contract is destroyed and re-deployed with different bytecode at the same address, each instance is considered a distinct version.

\begin{figure}[thbp]
    \centering
    \includegraphics[width=0.9\linewidth]{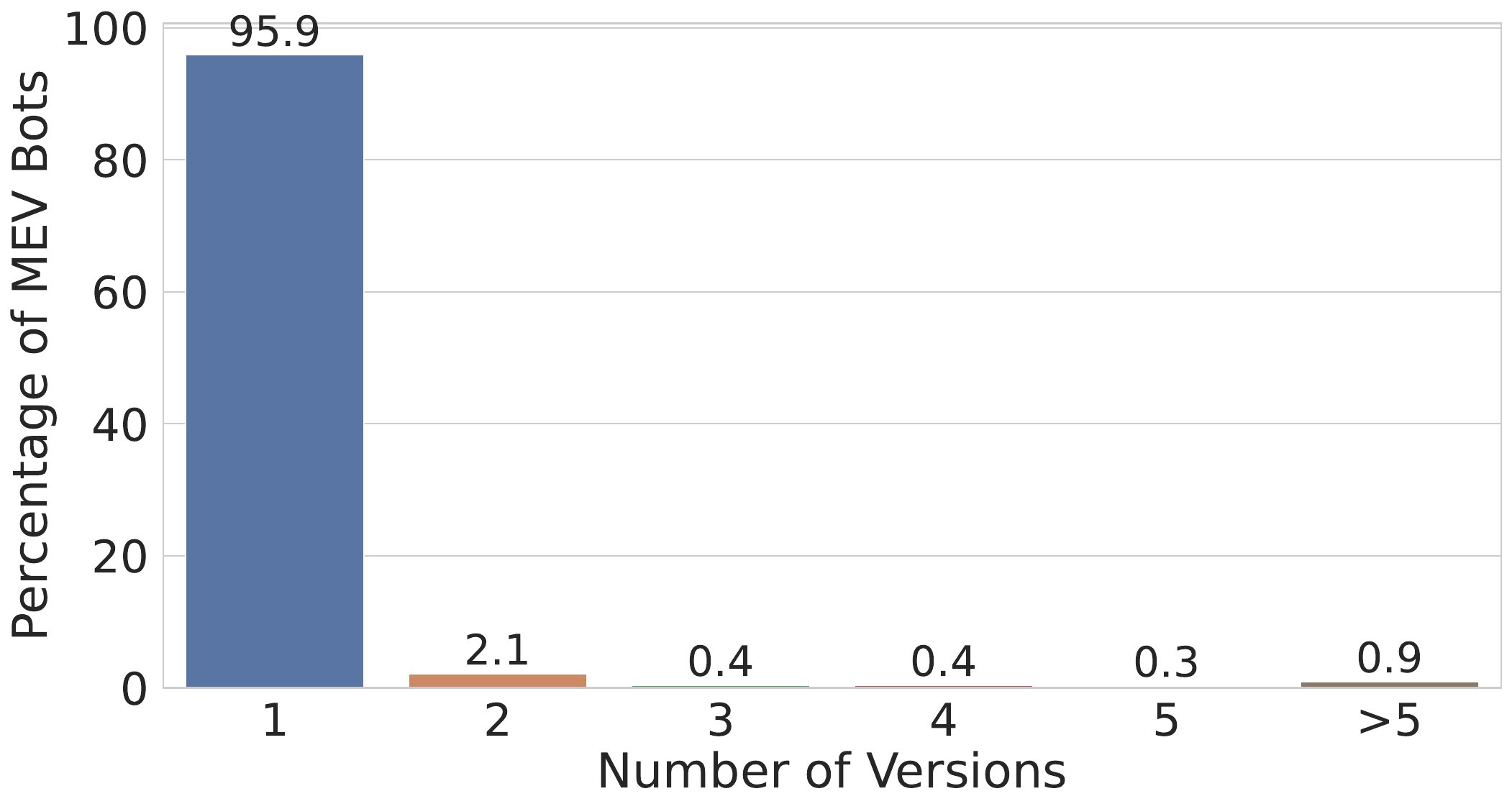}
    \caption{Distribution of the number of versions per MEV bot in our dataset.}
    \Description{A bar chart showing the distribution of version counts per MEV bot. The horizontal axis lists the number of versions as 1, 2, 3, 4, 5, and more than 5, while the vertical axis shows the percentage of MEV bots. The bar for one version dominates at 95.9 percent, while all other categories account for only small fractions.}
    \label{fig:hist-bots}
\end{figure}

\parhead{Version statistics of MEV bots}
To identify all versions of an MEV bot contract, we analyze the historical transactions associated with its address. Specifically, we parse transaction traces to check whether the address was the target of a \texttt{CREATE} or \texttt{CREATE2} deployment.
For each instance of bytecode deployment at the address, we record the corresponding version. This allows us to reconstruct the full sequence of code changes for MEV bot contracts that may have been redeployed multiple times.
We apply this method to all MEV bots in our dataset. As shown in~\autoref{fig:hist-bots}, about 96\% of them are deployed only once, indicating no code change over time. Meanwhile, about 0.9\% of MEV bots have more than five versions, suggesting that their code has been repeatedly updated.

\parhead{Vulnerability detection in historical versions}

For MEV bots with multiple historical versions, we also apply \name to each historical version to identify potential asset management vulnerabilities and attempt exploit generation. This analysis allows us to detect vulnerabilities that may exist only in earlier versions of the MEV bot contract but not in the current one.
In total, we identify 114 vulnerable historical versions across 44 MEV bot contracts and successfully generate 41 exploits targeting these versions.

\parhead{Loss estimation} To approximate the maximum possible loss, we analyze the highest ERC-20 token balance held by each exploitable version of an MEV bot. Specifically, for each exploitable version, we identify the block range during which that version was active and observe the maximum ERC-20 token balance within that range. If the version is the latest one, we consider the range from its deployment up to block 22,635,000 (June 5th, 2025).

Consistent with the estimation for current loss, if a contract's vulnerability is limited to a specific ERC-20 token, we consider only the potential loss associated with that token.
If an MEV bot has multiple exploitable versions, we report the loss corresponding to the version with the highest observed ERC-20 balance.
To simplify the estimation, we assume that all tokens were sold on June 5th, 2025. This allows us to use the token prices on Binance~\cite{binance2025datavision} on that day to estimate the potential loss for seven major tokens: WETH, WBTC, USDC, USDT, DAI, UNI and LINK.

\section{MEV Phishing Attacks}
\label{sec:additional-mev-phishing-attacks}

In this section, we provide more details on the remaining two variants of MEV phishing attacks: token-based and pool-based MEV phishing attacks.

\parhead{Token-based MEV phishing attacks}
A representative example of a \emph{token-based} MEV phishing attack was publicly documented in an early \href{https://x.com/bertcmiller/status/1421543838569705474}{case study} involving the \texttt{CHUM} token. In this incident, a searcher performed arbitrage on a newly deployed token that had been deliberately created as bait.

\begin{figure}
    \centering
    \includegraphics[width=0.85\linewidth]{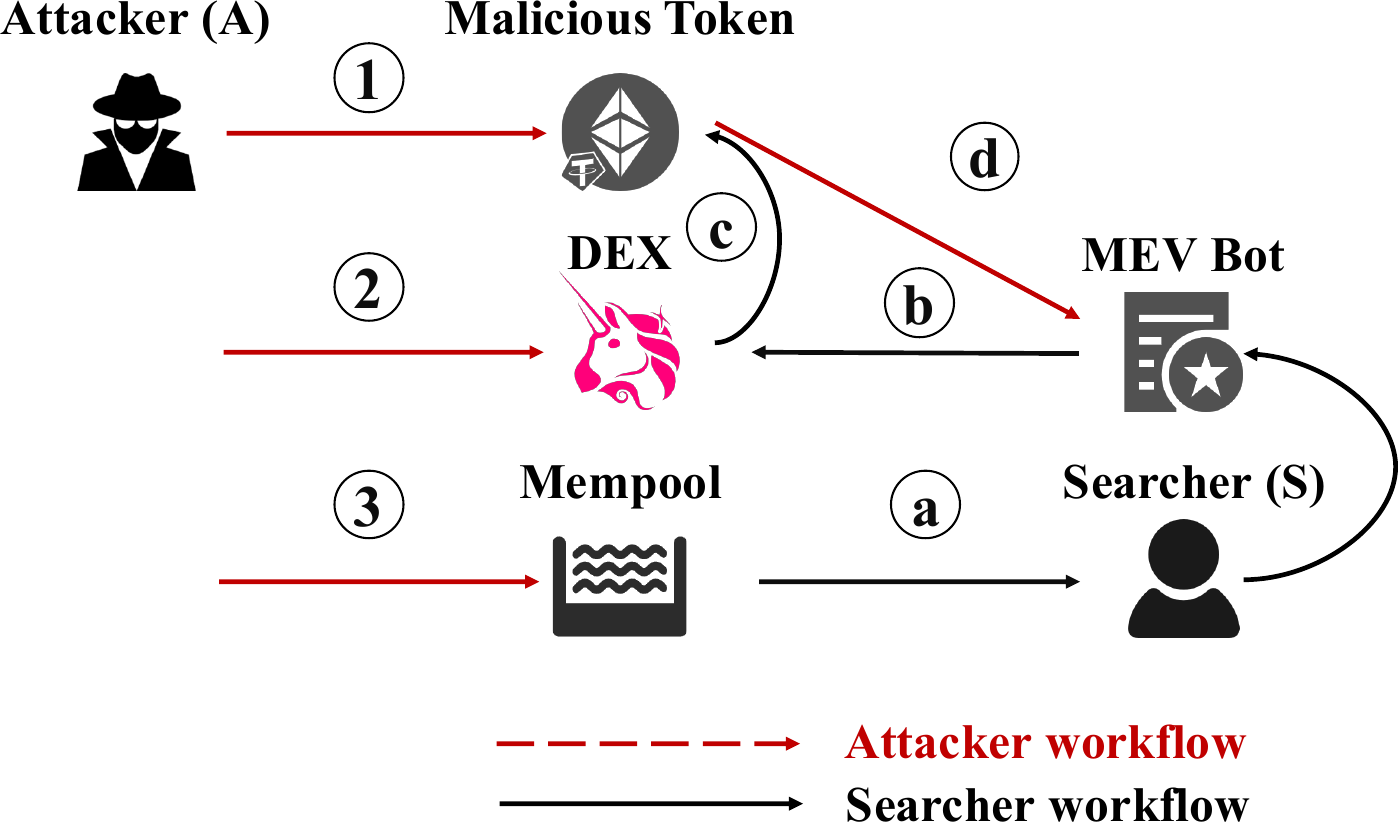}
    \caption{Token-based MEV phishing attack workflow.}
    \Description{A token-based MEV phishing attack workflow with two interacting paths. The attacker path, shown with red dashed arrows, prepares a malicious token, interacts with a decentralized exchange, and injects bait transactions into the mempool. The searcher path, shown with black solid arrows, observes the opportunity, interacts with the MEV bot, and triggers contract interactions involving the decentralized exchange and the malicious token.}
    \label{fig:token-based-mev-phishing}
\end{figure}

As shown in~\autoref{fig:token-based-mev-phishing}, the attack proceeds as follows. The attacker first deploys a malicious ERC-20 token contract (\textcircled{1}). Unlike a standard ERC-20 token implementation, the attacker modifies the \texttt{transfer} logic so that, when executed, it invokes a vulnerable function in a target MEV bot contract. This function is chosen to enable unauthorized asset transfers from the MEV bot contract.

Next, the attacker creates a liquidity pool for the malicious token, such as a Uniswap V2 pool (\textcircled{2}). After the pool is initialized, the attacker engineers an apparent MEV opportunity involving this pool, for example, by submitting a swap transaction. This transaction is broadcast to the public mempool, where it can be detected by MEV searchers monitoring for profitable opportunities (\textcircled{3}).

Upon observing the transaction, a searcher follows its standard workflow to extract the perceived MEV (\textcircled{a}). The searcher submits a transaction that invokes its MEV bot contract, which in turn executes a swap against the malicious pool (\textcircled{b}). During swap execution, the pool transfers the malicious token (\textcircled{c}), causing the token contract to be executed as part of the call chain.

At this point, the execution path includes the malicious token contract (searcher $\rightarrow$ MEV bot $\rightarrow$ pool $\rightarrow$ malicious token), while \texttt{tx.origin} remains the searcher.
The embedded malicious logic is then triggered (\textcircled{d}): the malicious token contract invokes a vulnerable function in the MEV bot contract, bypassing origin-based access control and transferring ERC-20 token assets from the bot to an attacker-controlled account.

\parhead{Pool-based MEV phishing attacks}
The second variant of MEV-phishing attacks is \emph{pool-based}, where the attacker deploys a malicious liquidity pool rather than a malicious token. In this setting, once a searcher executes a trade through the pool, the pool contract itself initiates an external call that exploits the MEV bot contract and drains its ERC-20 token balances.

A concrete instance of a pool-based MEV-phishing attack was observed in March 2025 (\href{https://etherscan.io/tx/0x1b539dbd898971f5296cea65c5af766df34f13dd7c1bea38e6700364847dc0d9}{0x1b5$\dots$0d9}). In this case, the attacker did not rely on a malicious token; instead, the attack was carried out entirely through a maliciously constructed pool contract.

\begin{figure}
    \centering
    \includegraphics[width=0.85\linewidth]{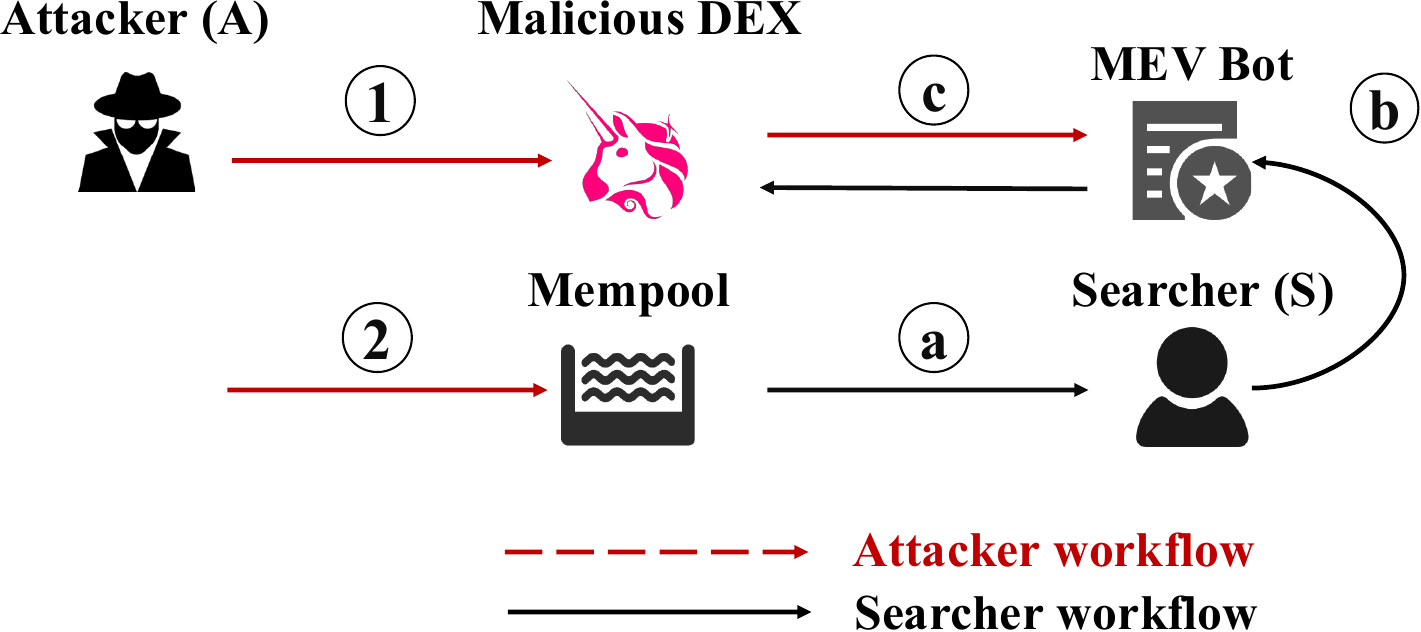}
    \caption{Pool-based MEV phishing attack workflow.}
    \Description{A pool-based MEV phishing attack workflow with two interacting paths. The attacker path, shown with red dashed arrows, deploys a malicious decentralized exchange and places bait transactions into the mempool. The searcher path, shown with black solid arrows, observes the opportunity, interacts with the MEV bot, and triggers execution involving the malicious exchange.}
    \label{fig:pool-based-mev-phishing}
\end{figure}

As shown in~\autoref{fig:pool-based-mev-phishing}, the attack proceeds as follows. The attacker first deploys a malicious pool contract whose constituent tokens are otherwise benign (\textcircled{1}). Similar to the token-based attack, the pool's swap logic is modified so that, upon execution of a swap, it invokes a vulnerable function in a target MEV bot contract.

As in other MEV-phishing attacks, the attacker then engineers an apparent MEV opportunity involving the malicious pool and broadcasts the corresponding transaction to the mempool (\textcircled{2}). A searcher observing this transaction follows its standard extraction workflow (\textcircled{a}) and submits a transaction that invokes its MEV bot contract (\textcircled{b}).
When the MEV bot executes the swap against the malicious pool, the pool's crafted logic is triggered (\textcircled{c}). The malicious pool contract becomes part of the execution path (searcher $\rightarrow$ MEV bot $\rightarrow$ malicious pool) and directly exploits the vulnerable function, transferring the bot's assets to the attacker.

\begin{figure}[htbp]
    \centering
    \begin{lstlisting}[language=Solidity, style=SolidityStyle, xleftmargin=5.0ex, breakindent=0pt]
contract UniswapV3PoolAccessControl {
    address FACTORY = 0x1F98431c8aD98523631AE4a59f267346ea31F984;
    bytes32 POOL_INIT_CODE_HASH = 
    0xe34f36d28c5efcd7c58e2e84af79e2a dffbe52f705d05dca7e6a181f8a19baf1;
    function verifyPoolAccess(
        address token0,
        address token1,
        uint24  fee
    ) public view returns (bool) {
        bytes32 salt = keccak256(abi.encode(token0,token1,fee));
        address computedAddress = address(uint160(uint256(
            keccak256(abi.encodePacked(
                bytes1(0xff),
                FACTORY,
                salt,
                POOL_INIT_CODE_HASH
            )))));
        return computedAddress == msg.sender;
    }
}
\end{lstlisting}
    \caption{A Solidity implementation to check if the caller is a valid Uniswap V3 pool. It verifies whether a given pool address is valid by deriving its expected address using \texttt{CREATE2}.}
    \Description{A Solidity implementation to check if the caller is a Uniswap V3 pool. It verifies whether a given pool address is valid by deriving its expected address using \texttt{CREATE2}.}
    \Description{A Solidity function named verifyPoolAccess that checks whether the caller is a valid Uniswap V3 pool. The code computes a deterministic pool address from token addresses, fee, the factory address, and the pool initialization code hash using CREATE2-related hashing, then compares the result with msg.sender.}
    \label{fig:check-pool-address}
\end{figure}

\section{Solidity Example of Checking Caller}
\label{sec:uniswap-v3-access-control}

\autoref{fig:check-pool-address} shows a Solidity example of how to rigorously check whether the caller is a Uniswap V3 pool.
The \texttt{verifyPoolAccess} function checks whether the caller is a valid Uniswap V3 pool by reconstructing the expected pool address using the \texttt{CREATE2} scheme. It first computes a salt by hashing the tuple \texttt{(token0, token1, fee)} with \texttt{keccak256}, consistent with how Uniswap V3 encodes pool parameters. Then, it derives the expected pool address using the standard \texttt{CREATE2} formula, which combines the factory address, the salt, and the pool's initialization code hash. If the derived address matches \texttt{msg.sender}, the function returns \texttt{true}, confirming that the caller is a legitimate pool contract created by the Uniswap V3 factory with the given parameters.

\end{document}